\documentclass[aps,prl,twocolumn,superscriptaddress]{revtex4-2}

\usepackage[hidelinks]{hyperref}
\usepackage{makecell}
\usepackage{physics}
\usepackage{balance}
\usepackage{suffix}
\usepackage{mathtools}
\usepackage[utf8]{inputenc}
\usepackage{booktabs}
\usepackage{cases}
\usepackage[multiple]{footmisc}
\usepackage{dcolumn}
\usepackage{color,soul}
\usepackage{rotating}
\usepackage{perpage}
\usepackage{siunitx}
\usepackage{xcolor}
\definecolor{darkcyan}{RGB}{0,139,139}
\hypersetup{
	colorlinks,
	linkcolor={red},
	citecolor={blue},
	urlcolor={blue}
}
\usepackage{soul}
\usepackage{amsmath}
\usepackage{tikz}
\usepackage[T1]{fontenc}
\usepackage{etoolbox}
\usepackage{graphics}
\usepackage{siunitx}
\usepackage{float}	
\usepackage{collref}
\usepackage{multirow}
\usepackage{mathtools}
\usepackage{bm}
\usepackage{url}
\usepackage{pifont}
\usepackage{balance}

\usepackage{amssymb}
\usepackage{mathtools}
\usepackage{amsmath}
\usepackage{bm}
\usepackage{siunitx}
\usepackage{graphicx}
\usepackage{float}
\usepackage{multirow}
\usepackage{booktabs}
\usepackage{tikz}
\usepackage{tikz-3dplot}
\usepackage{pgfplots}
\pgfplotsset{compat=1.18}

\usepackage{graphicx}
\usepackage{xcolor}
\usepackage{pifont}
\usepackage{diagbox}

\newcommand{\cmark}{\textcolor{green!60!black}{\huge\ding{51}}}
\newcommand{\xmark}{\textcolor{red}{\huge\ding{55}}}

\usepackage{tikz}
\usepackage{tikz-3dplot}
\usepackage{accents}

\makeatletter
\newcommand{\doublewidetilde}[1]{{%
		\mathpalette\double@widetilde{#1}%
}}
\newcommand{\double@widetilde}[2]{%
	\sbox\z@{$\m@th#1\widetilde{#2}$}%
	\ht\z@=.9\ht\z@
	\widetilde{\box\z@}%
}
\makeatother

\newcommand{\zb}{\color {black}}

\usepackage{etoolbox,lipsum}

\makeatletter
\let\aps@bib@device\bib@device
\renewcommand{\bib@device}[2]{%
	\begingroup
		\let\addcontentsline\@gobblethree
		\aps@bib@device{#1}{#2}%
	\endgroup
}
\makeatother

\newcommand{\blue}[1]{\textcolor{black}{#1}}

\begin{document} 
   \title{Parity-selective spin splitting in  coplanar antiferromagnets via bichromatic driving}
	\author{Di Zhu}
	\email{zhud33@mail2.sysu.edu.cn}
	\address{\mbox{Guangdong Provincial Key Laboratory of Magnetoelectric Physics and Devices, State Key Laboratory of} \mbox{Optoelectronic Materials and Technologies, School of Physics, Sun Yat-sen University, Guangzhou 510275, China}}
    \author{Zhongbo Yan}
    \email{yanzhb5@mail.sysu.edu.cn}
	\affiliation{\mbox{Guangdong Provincial Key Laboratory of Magnetoelectric Physics and Devices, State Key Laboratory of} \mbox{Optoelectronic Materials and Technologies, School of Physics, Sun Yat-sen University, Guangzhou 510275, China}}
	\author{Mohsen Yarmohammadi}
	\email{mohsen.yarmohammadi@georgetown.edu}
	\address{Department of Physics, Georgetown University, Washington DC 20057, USA}
	\date{\today}
	\begin{abstract}
		Parity is a central characteristic of momentum-dependent spin splitting in antiferromagnets (AFMs). Yet, 
intrinsic crystal symmetries typically restrict the splitting to either even or odd parity, preventing flexible spin control. Using a coplanar AFM, we demonstrate that bichromatic ($\omega$--$n\omega$) Floquet driving offers a natural way to bypass this constraint. 
This mechanism generates asymmetric spin textures unattainable in static AFMs or under monochromatic driving. For the specific AFM considered here, we reveal an elegant relation between spin-splitting parity and harmonic hierarchy: $\omega$--$2\omega$ fields generate highly tunable odd- and mixed-parity spin splittings, whereas higher-order harmonics ($n \ge 3$) exclusively produce even-parity states. The macroscopic magnetization can also be toggled via the specific driving protocol and harmonic $n$. These distinct spin-splitting states manifest in qualitatively different macroscopic spin currents generated after an optical quench---a definitive transport signature complementing direct visualization via spin- and angle-resolved photoemission spectroscopy.
	\end{abstract}
	
	\maketitle
	{\allowdisplaybreaks
		
		\blue{\textit{Introduction}}---The growing fascination with unconventional antiferromagnets (AFMs) is due to their distinctive momentum-dependent spin splitting (MDSS), which is centrally characterized by its parity~\cite{Hayami2019AM,LDYuan2020,LDYuan2021,Ma2021AM,libor2022AMa,Libor2022AMb,Libor2022AMc,song2025AM,Birk2023,Lin2025OAM,Yu2025pwave,zhuang2025AM9,Luo2026UAFM1,Luo2026UAFM2,Zeng2026OddAM}. Whether this splitting exhibits odd or even parity under momentum reversal ultimately dictates the material's unusual transport responses~\cite{Shao2021NC,Ouassou2023AM,Lu2024AM,Lin2024AM12,Zhuang2025SNL,Chakraborty2025AM,Brekke2024pwave,Hedayati2025pwave,Ezawa2025pwave,zhuang2026Mixed} as well as its propensity for hosting unconventional superconductivity~\cite{Zhu2023TSC,Brekke2023AM,Zhang2024AM,Chakraborty2024AM,Chakraborty2025AM2,Sun2025pwaveSC,Kim2026pwaveSC,Khodas2026pwaveSC,Lee2024dwaveNAFM,Zhang2025AMSC,zhang2026CAFM}. 
        
        While these well-defined symmetry features have made AFMs promising platforms for diverse applications~\cite{RevModPhys.90.015005,Jungwirth2016,PhysRevLett.126.127701}, the fixed parity of spin splitting imposed by the intrinsic symmetries of the crystal lattice~\cite{Liu2022AM,Xiao2024SSG,Yi2024SSG,Chen2023AM} also restricts the range of accessible physical phenomena and the applicability of any given material. Achieving on-demand spin control thus requires a means to overcome this limitation. This naturally raises the question: Is there a universal route to achieving controllable MDSS parity in AFMs?
        
        Floquet engineering ~\cite{nn5t-kmln,7ywb-ml2q,9346-9jpf,k3xb-8pts,xt23-9pnv,zou2026floquetengineeredoddparityaltermagnetichigherorder,yarmohammadi2026floquetinducedanisotropicmagnetoresistanceanomalous,yarmohammadi2026giantperpendicularedelsteinpolarization,yarmohammadi2026efficienttwocolorfloquetcontrol,yu2026tunableoddparityspinsplittings} has recently been shown to be a highly flexible route for generating unconventional spin textures that are symmetry-forbidden in pristine AFMs~\cite{9346-9jpf,Li2025AM7,7ywb-ml2q,liu2025AM10,nn5t-kmln,liu2026AM,zhu2026BCAFM}. Yet the predominant monochromatic driving scheme offers only a partial circumvention of lattice symmetry restrictions; as a result, the parity of the induced MDSS remains limited to a predetermined form. Microscopically, light-driven band modification arises from the absorption and emission of virtual photons. 
        Compared with single-color (frequency) driving, a multi-color driving field activates a richer array of such processes~\cite{yarmohammadi2026floquetinducedanisotropicmagnetoresistanceanomalous,yarmohammadi2026giantperpendicularedelsteinpolarization,yarmohammadi2026efficienttwocolorfloquetcontrol,yarmohammadi2026chirpedfloquetlineardrives,yu2026tunableoddparityspinsplittings}, thereby affording greater flexibility in band-structure engineering. This reasoning points to multi-color driving as a promising strategy for overcoming the parity bottleneck of monochromatic schemes.\begin{table}[t]
			\resizebox{1\linewidth}{!}{
				\begin{tabular}{c|c|c|c}
					\hline\hline
					\diagbox[width=2cm, height=1cm]{Spin \\ splitting}{System} & 
					\begin{tabular}[c]{@{}c@{}} Pristine BCAFM \end{tabular} & 
					\begin{tabular}[c]{@{}c@{}} Monochromatically \\ driven BCAFM\end{tabular} & 
					\begin{tabular}[c]{@{}c@{}} Bichromatically \\ driven BCAFM\end{tabular} \\
					\hline
					\begin{tabular}[c]{@{}c@{}} Odd-parity\\\end{tabular} & \xmark & \xmark & \cmark \\
					
					\begin{tabular}[c]{@{}c@{}} Even-parity\\\end{tabular} & \xmark & \cmark & \cmark \\
					
					\begin{tabular}[c]{@{}c@{}} Mixed-parity\\\end{tabular}  & \xmark & \xmark & \cmark \\
					\hline\hline
			\end{tabular}}\caption{\textbf{Symmetry-allowed spin-splitting parities.} Summary of allowed spin-splitting parities in our bilayer coplanar AFM~(BCAFM) under varying driving field configurations. In the absence of a driving field, the system remains entirely spin-degenerate. While conventional monochromatic driving breaks sufficient symmetries to allow even-parity spin splitting, it preserves the symmetries forbidding odd-parity terms. In contrast, bichromatic driving lifts these constraints, uniquely enabling asymmetric odd- and mixed-parity spin splittings.}				\label{tab1}
		\end{table}

        In this Letter, we demonstrate that bichromatic ($\omega$--$n\omega$) driving---the simplest extension of monochromatic driving---provides deterministic parity control of MDSS in a bilayer coplanar AFM (BCAFM) [Table~\ref{tab1}]. By tailoring the two-color optical field, we 
        selectively preserve certain symmetries to restrict the induced spin polarization to be unidirectional and out-of-plane. Our systematic investigation reveals a harmonic hierarchy with and without net magnetization~($s$-wave) depending on the driving protocol and harmonic $n$: $\omega$--$2\omega$ fields dynamically generate highly tunable odd-~($p$-wave) and mixed-parity~($p+d$- and $s+p+d$-wave)  MDSSs, whereas higher-order harmonics ($n \ge 3$) exclusively produce even-parity ($s+d$- and $d$-wave) states. We further demonstrate that even- and odd-parity components can be unequivocally distinguished via quench-induced macroscopic spin currents, providing a transport probe of the parity control that complements direct visualization through spin-resolved angle-resolved photoemission spectroscopy~(ARPES).

		\begin{figure}[t]
        \centering
        \includegraphics[width=0.95\linewidth]{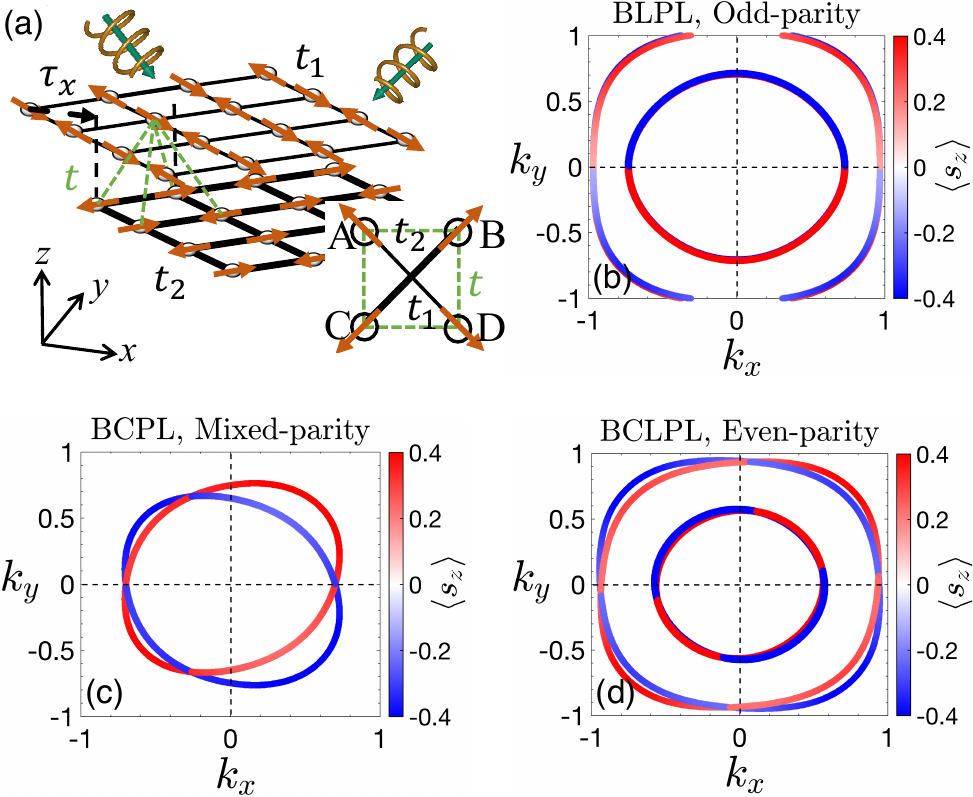}
        \caption{\textbf{Dynamically generated spin-splitting parities.} (a) Schematic of the all-out BCAFM under incident bichromatic light. The system consists of two collinear square-lattice monolayers offset by $\bm{\tau}_x$, featuring mutually orthogonal N\'eel vectors, intralayer hoppings $t_1$ and $t_2$, and interlayer hopping $t$. (b)-(d) Momentum-resolved spin polarization $\langle s_z \rangle$ in the $k_x$-$k_y$ plane for $n=2$ ($\omega$--$2\omega$) bichromatic driving. The driving fields correspond to (b) BLPL~($n = 2$, $\beta = \pi/4$ and $\phi = \pi/3$), (c) BCPL~($n = 2$, $\phi = 0$ and $\eta_1 = \eta_2 = +1$), and (d) BCLPL~($n = 3$, $\beta = \pi/4$, $\phi = \pi/3$, and $\eta_1 = +1$). BLPL induces a strict odd-parity spin texture, whereas BCPL and BCLPL generate mixed-parity and even-parity spin splittings, respectively (see Sec.~S1 of the SM~\cite{SM} and Table~\ref{tab2} for special conditions under which the BCPL and BCLPL cases also show purely odd and mixed parities and complete cancellation). Unless otherwise specified, the system parameters are set to $\mu\simeq 0.34$ and $(t, t_s, t_a, M_x,M_y) = (0.4, 0.7, 0.3, 0.5, 0.5)$, with driving parameters $(A_0, S, \omega) = (0.6, 1, 5)$.}\label{fig1_new}
        \end{figure}

		\blue{\textit{Bilayer coplanar AFM with spin degeneracy}}---The BCAFM for study is sketched in Fig.~\ref{fig1_new}(a).
        It comprises two collinear AFM square monolayers offset by $\bm{\tau}_x=a(1,0)$.
        Sharing a lattice constant $\sqrt{2}a$, their mutually perpendicular Néel vectors establish an all-out coplanar magnetic configuration. 
        Such a BCAFM can be realized in engineered two-dimensional heterostructures~\cite{hejazi2020noncollinear,robertson2025imaging,LiuTwistAM2024,BaiTypeIVAFM2025,9346-9jpf}. Specifically, stacking square-lattice van der Waals AFMs with an exact $90^{\circ}$ twist successfully enforces the requisite orthogonal spin textures. Alternatively, precision-grown transition-metal oxide superlattices (e.g., LaMnO$_3$/LaNiO$_3$~\cite{gibert2015interfacial}) can robustly stabilize these magnetic configurations via interfacial coupling.

            The tight-binding Hamiltonian for this BCAFM is given by $
            H=\sum_{\langle i,j\rangle,m\neq n,\sigma}tc^\dagger_{i,m,\sigma}c_{j,n,\sigma}+\sum_{\langle i,j\rangle,m,\sigma}t_{m}c^\dagger_{i,m,\sigma}c_{j,m,\sigma}+\sum_{i,m,\sigma,\sigma'}c^\dagger_{i,m,\sigma}(\bm{M}_{i,m}\cdot \bm{s})_{\sigma,\sigma'}c_{i,m,\sigma'}$, where $c_{i,m,\sigma} (c_{i,m,\sigma}^\dagger)$ is the electron annihilation (creation) operator at site $i$, layer $m = 1,2$, and spin $\sigma$~\cite{zhu2026BCAFM}. The three terms represent, respectively, interlayer hopping ($t$), layer-dependent intralayer hopping ($t_m$), and the exchange interaction with local magnetic moments $\bm{M}_{i,m}=(\pm M_x,\pm M_y)$, with $\bm{s}=(s_x,s_y)$ denoting the Pauli spin vector. Transforming to the basis $\psi_{\bm{k}}=(c_{\bm{k},\uparrow},c_{\bm{k},\downarrow})^{T}$ with $c_{\bm{k},\sigma}=(c_{A,\bm{k},\sigma},\dots,c_{D,\bm{k},\sigma})^{T}$, the momentum-space Hamiltonian is given by \cite{zhu2026BCAFM}
            \begin{align}
\mathcal{H}(\bm{k})={} &2t h^{c}_{x}\sigma_x+2t h^{c}_{y}\tau_x+4t_{s}h^{c}_{x}h^{c}_{y}\tau_x\sigma_x\notag\\
&+4t_{a}h^{c}_{x}h^{c}_{y}\tau_y\sigma_y-M_x\sigma_zs_x+M_y\tau_zs_y\,.
\end{align} Here, $h^{c}_{j}\equiv\cos k_{j}$ ($j=x,y$) and identity matrices are omitted. The Pauli matrices follow the tensor order $\tau\otimes\sigma\otimes s$ spanning the four sublattice and two spin degrees of freedom. Intralayer hoppings are decomposed into symmetric $t_{s}=(t_1+t_2)/2$ and antisymmetric $t_{a}=(t_2-t_1)/2$ parts. Throughout, we set $a = 1$ and take $t, t_s, t_a > 0$. Without driving, the $\mathcal{PT}$-like $[\bar{E}\|\mathcal{T}C_{2z}]$ symmetry enforces Kramers degeneracy, yielding a spin-degenerate spectrum with zero net spin polarization at all momenta. Upon application of monochromatic circular driving, the system has been shown to develop symmetry-enforced 
even-parity ($d$-wave) spin splitting~\cite{zhu2026BCAFM}.
 
        \blue{\textit{Bichromatic driving and parity-selective spin splitting}}---To elucidate the effects of bichromatic driving, we expand the Hamiltonian around the $\Gamma$ point. This region dictates the dominant vertical optical transitions probed by experiments~\cite{https://doi.org/10.1002/adma.201302616,PhysRevLett.108.117403,RevModPhys.93.025006,Basov2017,Hajlaoui2014}. Substituting $h_j^c \approx 1-k_j^2/2$ and keeping terms up to $\mathcal{O}(k^2)$, we obtain 
        
        {\small\begin{align}\label{ConModel}
        \mathcal{H}(\bm{k})\approx{}&t(2-k_x^2)\sigma_x+t (2-k_y^2)\tau_x+t_{s}(4-2k_x^2-2k_y^2)\tau_x\sigma_x\notag\\
        &+t_{a}(4-2k_x^2-2k_y^2)\tau_y\sigma_y-M_x\sigma_zs_x+M_y\tau_zs_y.
        \end{align}}

       To incorporate the $\omega$--$n\omega$ bichromatic driving, we apply the Peierls substitution $\bm{k}\rightarrow \bm{k}+\bm{A}(t)$ with $e=\hbar=1$. The time-periodic Hamiltonian is then Fourier expanded as $\mathcal{H}(\bm{k}+\bm{A}(t))=\sum_{m\in\mathbb{Z}}\mathcal{H}_m e^{im\omega t}$. In the high-frequency off-resonant regime, the system is described by the effective static Floquet Hamiltonian \cite{Kitagawa2011EffectiveH,Goldman2014EffectiveH,eckardt2015EffectiveH}
        \begin{align}
        \mathcal{H}_{\rm eff}(\bm{k})=\mathcal{H}_0(\bm{k})+\sum_{m\ge1}\frac{[\mathcal{H}_m,\mathcal{H}_{-m}]}{m\omega}+\mathcal{O}(\omega^{-2})\,.
        \end{align}
        To analyze the driving-induced spin splitting, we initially focus on the $n=2$ ($\omega$--$2\omega$) case, with generalizations to higher-order harmonics ($n \geq 3$) explored in subsequent sections. For $n=2$, the general vector potential is $\bm{A}(t)=A_0\big(\cos\omega t+S\cos(2\omega t+\phi)\cos\beta,\, \eta_1\sin{\omega t}+\eta_2S\cos{\small(}2\omega t+\phi-\tilde{\phi}{\small)}\cos{\small(}\beta-\tilde{\beta}{\small)}\big)$, where $S$ is the relative amplitude of the $2\omega$ field, $\eta_{1,2}$ define the polarization helicities, $\phi, \tilde{\phi}$ are relative phases, and $\beta, \tilde{\beta}$ are polarization angles. By tuning these parameters, we investigate three configurations: bichromatic linearly (BLPL), circularly (BCPL), and circular-linear (BCLPL) polarized light. The resulting effective Hamiltonians take the form
        \begin{align}
         \mathcal{H}_{\rm eff}^{\alpha}(\bm{k})={}&2tX^{\alpha}\sigma_x+2tY^{\alpha}\tau_x+4t_sR^{\alpha}\tau_x\sigma_x+4t_aR^{\alpha}\tau_y\sigma_y\notag\\
         &-M_x\sigma_zs_x+M_y\tau_zs_y+F^{\alpha}\tau_y\sigma_z-F^{\alpha}\tau_z\sigma_y\,,
         \end{align}where $\alpha \in \{{\rm BLPL, BCPL, BCLPL}\}$. Here, $\{X^\alpha, Y^\alpha, R^\alpha\}$ represent rescalings of the bare hopping parameters, while the light field dynamically generates the $F^\alpha$ terms. The generated imaginary hopping phase ($\tau_y \sigma_z$, $\tau_{z}\sigma_{y}$) by the bichromatic field acts as a gauge flux, inducing a directional flow pattern on the lattice that breaks the spin-crystal symmetries responsible for enforcing spin degeneracy~\cite{zhu2026BCAFM}. Physically, the interplay between this light-induced directional flow and the intrinsic coplanar magnetic exchange fields inherently couples the electron's spin to its momentum, giving rise to MDSS whose symmetry pattern is set entirely by the form factors of the $F^\alpha$ terms (see Sec.~S2 of the SM~\cite{SM} for an analytical demonstration based on the high-temperature-expansion method~\cite{Hayami2020AM1,Hayami2020AM2}). 

        The BLPL configuration corresponds to $\eta_1=0$, $\eta_2=1$, $\tilde{\phi}=0$, and $\tilde{\beta}=\frac{\pi}{2}$, yielding $\bm{A}(t)=A_0\big(\cos\omega t+S\cos(2\omega t+\phi)\cos\beta,\, S\cos(2\omega t+\phi)\sin\beta\big)$. Truncating at $\mathcal{H}_{\pm 2}$, we obtain: $X^{\rm BLPL}=1-k_x^2/2-A_0^2(1+S^2\cos^2\beta)/4$, $Y^{\rm BLPL}=1-k_y^2/2-A_0^2S^2\sin^2\beta/4$, $R^{\rm BLPL}=X^{\rm BLPL}+Y^{\rm BLPL}-1$, and $F^{\rm BLPL}=\frac{tt_aA_0^3S}{\omega}k_y\sin\beta\sin\phi$~(see Sec.~S1 of the Supplemental Materials~(SM)~\cite{SM}). Because $F^{\rm BLPL}$ is proportional to $k_y$ (for $\sin\beta\sin\phi\neq0$), the BLPL induces a $p$-wave spin splitting [Fig.~\ref{fig1_new}(b)]. Crucially, this term depends linearly on $S$, which implies that the $p$-wave spin splitting is strictly a two-beam effect. From a symmetry perspective, single-color LPL preserves inversion and effective $\mathcal{PT}$ symmetries, leaving the bands spin-degenerate. Introducing a secondary $2\omega$ LPL field breaks both symmetries while preserving effective time-reversal symmetry, ultimately lifting the degeneracy to yield an odd-parity spin splitting.

        The BCPL case ($\eta_1,\eta_2\neq 0$, $\tilde{\phi}=\frac{\pi}{2}$, $\beta=\tilde{\beta}=0$) yields $\bm{A}(t)=A_0\big(\cos(\omega t) + S \cos(2\omega t + \phi),\, \eta_1 \sin(\omega t) + \eta_2 S \sin(2\omega t + \phi)\big)$. Truncating at $\mathcal{H}_{\pm 2}$, we find: $X^{\rm BCPL}=1-k_x^2/2-(1+S^2)A_0^2/4$, $Y^{\rm BCPL}=1-k_y^2/2-(1+S^2)A_0^2/4$, $R^{\rm BCPL}=X^{\rm BCPL}+Y^{\rm BCPL}-1$, and $F^{\rm BCPL}=\frac{tt_aA_0^2}{\omega}\big[(4\eta_2 \eta_1+1)A_0Sk_x\sin\phi-(4\eta_1+\eta_2)A_0Sk_y\cos\phi-(8\eta_1+4\eta_2 S^2)k_x k_y\big]$. The dynamically generated $F^{\rm BCPL}$ term introduces both $d$-wave and $p$-wave components. The $d$-wave term vanishes completely for counter-rotating beams ($\eta_1=-\eta_2$) with $S^2=2$, while the persistent $p$-wave terms can be tuned to purely $p_y$ or $p_x$ by setting $\sin\phi=0$ or $\cos\phi=0$. Ultimately, the BCPL driving allows precise tuning between strict $p$-wave and mixed-parity $(p+d)$-wave [Fig.~\ref{fig1_new}(c)] spin splittings. This result also admits a symmetry-based interpretation: introducing the $2\omega$ harmonic breaks inversion symmetry, which, combined with the time-reversal symmetry breaking of the fundamental CPL, 
        admits the generic coexistence of even- and odd-parity components.

        The BCLPL case ($\eta_1\neq 0$, $\eta_2 = 1$, $\tilde{\phi}=0$, $\tilde{\beta}=\frac{\pi}{2}$) gives $\bm{A}(t)=A_0\big(\cos\omega t+S\cos(2\omega t+\phi)\cos\beta,\, \eta_1\sin\omega t+ S\cos(2\omega t+\phi)\sin\beta\big)$. Truncating at $\mathcal{H}_{\pm 3}$ yields  $X^{\rm BCLPL}=1-k_x^2/2-A_0^2(1+S^2\cos^2\beta)/4$, $Y^{\rm BCLPL}=1-k_y^2/2-A_0^2(1+S^2\sin^2\beta)/4$, $R^{\rm BCLPL}=X^{\rm BCLPL}+Y^{\rm BCLPL}-1$, and $F^{\rm BCLPL}=\frac{tt_aA_0^2}{\omega}\big[-8\eta_1 k_x k_y+A_0S(\sin\beta\sin\phi-4\eta_1\cos\beta\cos\phi)k_y+A_0S(\cos\beta\sin\phi+4\eta_1\sin\beta\cos\phi)k_x+4A_0^2S^2\eta_1\cos\beta\sin\beta/3\big]$. This $F^{\rm BCLPL}$ term dictates that the $d$-wave ($k_x k_y$) component is always present [Fig.~\ref{fig1_new}(d)]. The $p_y$ or $p_x$ components vanish if $\tan\beta\tan\phi=4\eta_1$ or $\tan\phi=-4\eta_1\tan\beta$, respectively, but never simultaneously since $\tan^2\beta=-1$ has no real solution. Alongside an $s$-wave term (active when $\sin(2\beta)\neq 0$), the linear harmonic transforms the pure $d$-wave CPL response into a mixed-parity ($s+p+d$-wave) state capable of hosting net magnetization. As in the BCPL case, combining the fundamental CPL with a $2\omega$ driving field breaks both time-reversal and inversion symmetries, giving rise to a mixed-parity spin texture. The middle column of Table~\ref{tab2} provides a summary of these three cases.
        
        \begin{table}[t]
			\renewcommand{\arraystretch}{1.4} 
			\begin{ruledtabular}
				\begin{tabular}{ccc}
					\makecell{Driving \\field} & \makecell{Harmonic \\ $n=2$} & \makecell{Harmonic \\$n\geq 3$} \\
					\colrule
					BLPL  & Odd   & None \\
					BCPL  & Mixed\footnotemark[1] & Even\footnotemark[2]  \\
					BCLPL & Mixed & Even \\
				\end{tabular}
                \footnotetext[1]{Reduces to Odd if $S^2=2$ and $\eta_1 = -\eta_2$.}
			\footnotetext[2]{Reduces to None if $S^2=n$ and $\eta_1 = -\eta_2$.}
                \caption{\textbf{Floquet parities under higher-harmonic bichromatic driving.} Summary of the spin-splitting parity for the continuous model in Eq.~\eqref{ConModel} under different two-color driving fields by various higher-order harmonics $n$, determined by their momentum ($\bm{k}$) dependence. Footnotes indicate specific parameter tunings where counter-rotating beams ($\eta_1 = -\eta_2$) cause the generated terms to undergo parity reduction or complete cancellation (detailed analytical expressions are provided in the Sec.~S3 of the SM~\cite{SM}).}\label{tab2}
			\end{ruledtabular}	
\end{table}

        \blue{\textit{Generalization to higher-order harmonics ($n\geq3$)}}---Extending our framework to higher-order harmonics (derivations are provided in Sec.~S3 of the SM~\cite{SM}), we find that the band structure either retains spin degeneracy or exhibits an exclusively even-parity spin splitting (see the right column of Table~\ref{tab2}). For $n\geq3$, purely linear driving (BLPL) can no longer generate the $F^{\alpha}$ terms, so the spin degeneracy persists (see Table S1 in Sec.~S3 of the SM~\cite{SM}). For purely circular driving (BCPL), the $F^{\alpha}$ terms become an even function of momentum, 
        leading exclusively to even-parity spin splitting. However, unlike the monochromatic CPL case, which produces a purely 
        $d$-wave spin splitting~\cite{zhu2026BCAFM}, the two-color BCPL with $n=3$ generally yields a mixture of $s$- and $d$-wave components (see Table S2 in Sec.~S3 of the SM~\cite{SM}), allowing for a finite magnetization. For $n\geq4$, although the spin splitting reduces to a purely $d$-wave form, a notable distinction remains: 
        the $d$-wave spin splitting can be completely suppressed when the two beams are counter-rotating ($\eta_1 = -\eta_2$) with matched relative intensities ($S^2 = n$). In the mixed (BCLPL) driving case,  the spin splitting is generally a mixture of $s$- and $d$-wave components for $n\geq3$ (see Table S3 in Sec.~S3 of the SM~\cite{SM}). Interestingly, while the $d$-wave is always present, the $s$-wave components can be tuned to vanish by adjusting the parameters of the two beams.
        
        In what follows, we demonstrate that these unconventional two-color light-induced spin splittings can be unambiguously probed either through macroscopic spin transport generated by a sudden optical quench or directly via spin-resolved ARPES.\begin{figure}[t]
        	\centering
        	\includegraphics[width=1\linewidth]{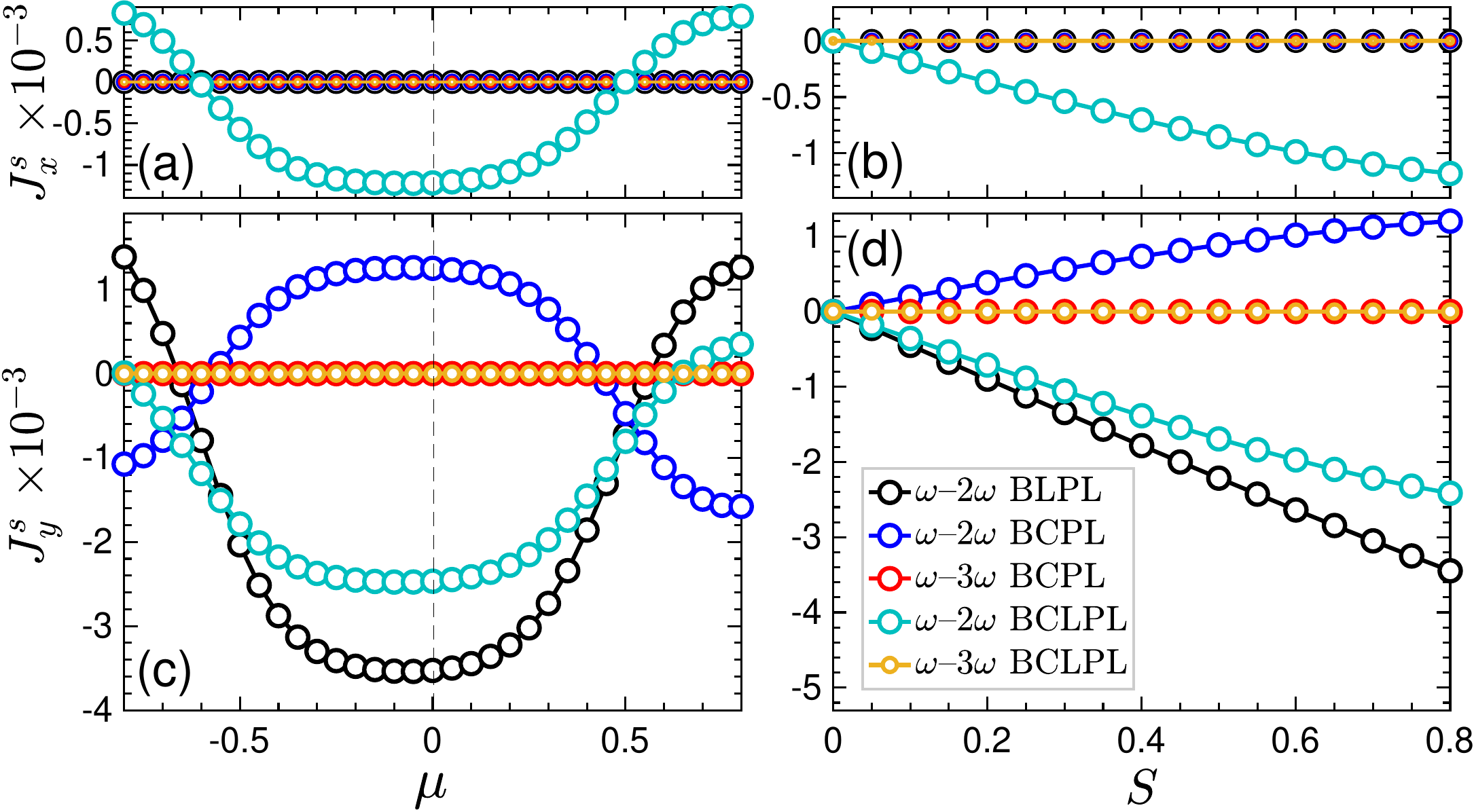}
        	\caption{\textbf{Macroscopic spin current response following a sudden optical quench.} Spin currents $J^s_x$ and $J^s_y$ as a function of (a,c) the chemical potential $\mu$ ($A_0=0.6$ and $S=0.8$) and (b,d) the secondary field strength $S$ ($A_0=0.6$ and $\mu=0.1$). Common parameters are $t=0.4$, $t_s=0.7$, $t_a=0.3$, $\omega=10$, $M_x=M_y=0.5$, and $T=0.1$. Specific configurations: BLPL ($n=2$, $\phi=\frac{\pi}{2}$, and $\beta=\frac{\pi}{2}$); BCPL ($n \in \{2,3\}$, $\eta_1=1$, $\eta_2=1$, and $\phi=0$); BCLPL ($n \in \{2,3\}$, $\eta_1=1$, $\phi=\frac{\pi}{3}$, and $\beta=\frac{\pi}{4}$). Only $n=2$ driving successfully generates finite spin currents.}\label{fig2_new}
        \end{figure}
        
            \blue{\textit{Spin current for a quench
        protocol}}---We first propose using the spin current generated after an optical quench to probe the light-induced spin splittings. Assuming no irradiation at $t \le 0$, turning on the optical field projects the initial states $|\psi_{{\rm in},\bm{k}}\rangle$ in the tight-binding model onto the Floquet eigenstates $|\psi_{n,\bm{k}}\rangle$ \cite{Dehghani2015quench}. The corresponding occupation probabilities, $\rho^{\rm quench}(\bm{k})=|\langle\psi_{{\rm in},\bm{k}}|\psi_{n,\bm{k}}\rangle|^2$, define the directional spin current as
        \begin{align}
            J_{x,y}^s=-\int dk^2  \rho^{\rm quench}(k) v_{x,y}^s(k)\,,
        \end{align}where $v_{x,y}^s(\bm{k})=\frac{1}{2}\{s_z, \partial_{k_{x,y}} \mathcal{H}_{\rm eff}\}$ is the spin velocity operator. Because a macroscopic spin current necessitates broken inversion symmetry, it is exclusively driven by the odd-parity components generated by the secondary optical field.

            In Fig.~\ref{fig2_new}, we calculate the spin currents $J^s_x$ and $J^s_y$ generated by the odd-parity $p_x$ and $p_y$ spin-splitting components. Because the $\omega$--$3\omega$ BCPL and BCLPL fields strictly generate even-parity states (Table~\ref{tab2}), they preserve spatial inversion symmetry and thus cannot drive a macroscopic spin current [Figs.~\ref{fig2_new}(a) and \ref{fig2_new}(c)]. In contrast, the $\omega$--$2\omega$ fields induce the necessary odd-parity terms. Specifically, the $\omega$--$2\omega$ BLPL and BCPL fields generate only a $p_y$ component (producing $J^s_y$), while the BCLPL field generates both $p_x$ and $p_y$ components (producing both $J^s_x$ and $J^s_y$). As shown in Figs.~\ref{fig2_new}(b) and \ref{fig2_new}(d), the magnitude of these spin currents scales directly with the strength of the second light field $S$. Ultimately, since odd-parity spin splitting emerges exclusively via the second field, the spin current provides a highly sensitive probe for two-color Floquet dynamics.\begin{figure}[t]
		\centering
		\includegraphics[width=0.8\linewidth]{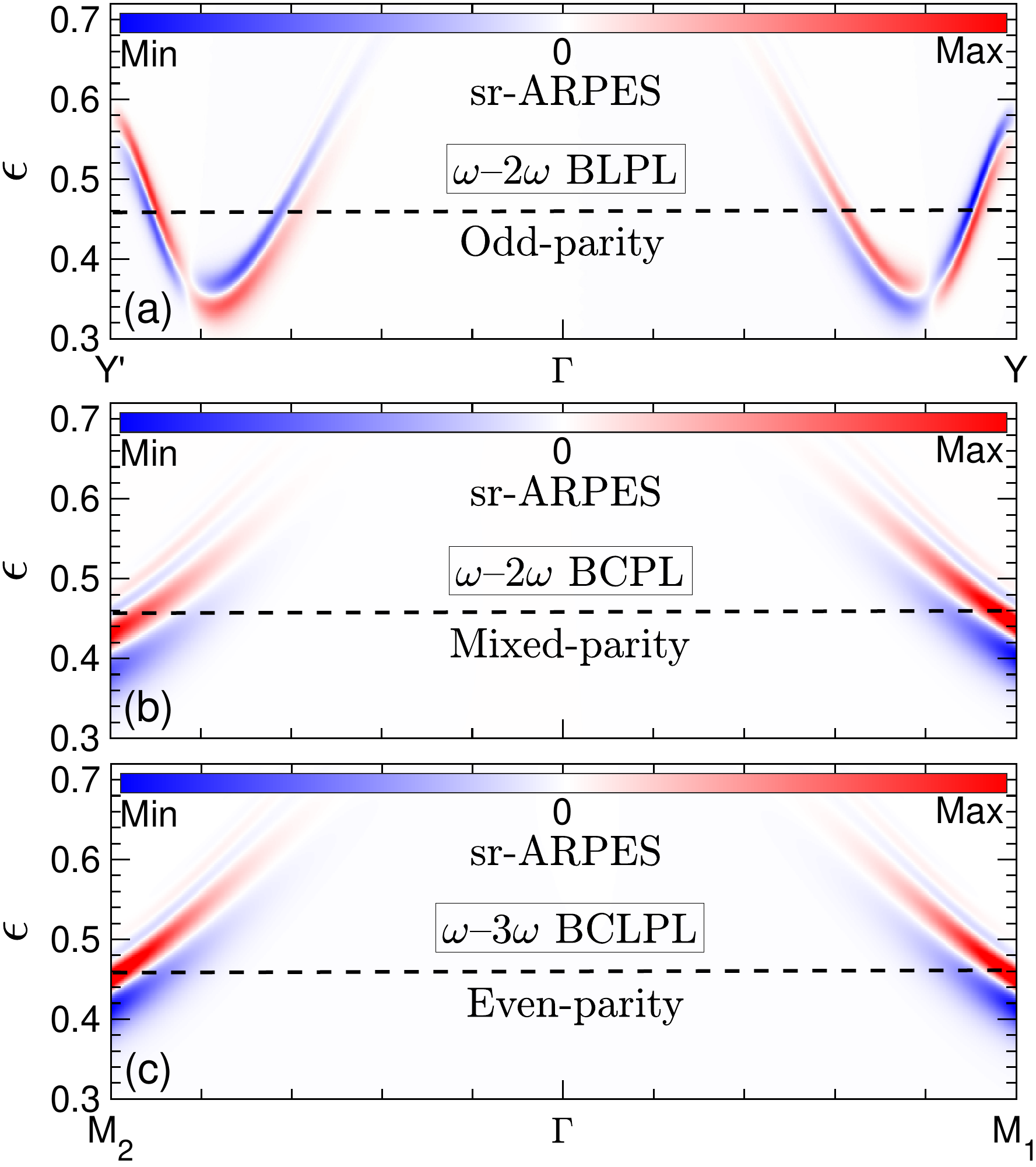}
		\caption{\textbf{Simulated spin-resolved ARPES (sr-ARPES) spectra.} (a) The $n=2$ BLPL case evaluated along $Y'[(0,\frac{\pi}{2})]$--$\Gamma$--$Y[(0,-\frac{\pi}{2})]$, indicating odd-parity spin splitting ($S=0.8$, $\phi=\frac{\pi}{2}$, $\beta=\frac{\pi}{2}$). (b,c) Spectra along $M_2[(-\frac{\pi}{4},-\frac{\pi}{4})]$--$\Gamma$--$M_1[(\frac{\pi}{4},\frac{\pi}{4})]$. (b) The $n=2$ ($\omega$--$2\omega$) BCPL driving yields a mixed-parity texture ($S=0.6$, $\phi=0$, $\eta_1= \eta_2 = 1$), whereas (c) the $n=3$ ($\omega$--$3\omega$) BCLPL driving strictly enforces even parity ($S=0.6$, $\eta_1=1$, $\phi=\frac{\pi}{3}$, $\beta=\frac{\pi}{4}$). Parameters: $t=0.4$, $t_s=0.7$, $t_a=0.3$, $A_0=0.6$, $\omega=6$, $M=0.5$, $\mu=0.5$, $t_d=20$, $T=0.1$, $T_{\rm probe}=80$, and $T_{\rm pump}=100$.}\label{fig3_new}
		\end{figure}
            
			\blue{\textit{Spin-resolved ARPES}}---As the second method to experimentally probe these tunable spin splittings, we introduce spin-resolved ARPES~\cite{doi:10.1126/science.1167733,Iwasawa2024}. In a standard pump-probe setup, we assume Gaussian envelopes for the pump and probe pulses through{\small\begin{align}
			\bm{A}_{\rm pump}(t)=e^{-\frac{t^2}{2T^2_{\rm pump}}}\bm{A}(t),\quad s(t,t_d)=e^{-\frac{(t-t_d)^2}{2T^2_{\rm probe}}}\,.
		\end{align}}Here, $\bm{A}(t)$ is the continuous two-color driving field employed in the present work, $s(t,t_d)$ is the temporal profile of the incident probe laser pulse, $T_{\rm pump, probe}$ dictate the respective pulse widths, and $t_d$ is the delay time relative to the pump peak at $t=0$. The time-resolved ARPES photocurrent intensity is~\cite{7ywb-ml2q,SentefTrARPES,FreeeicksTrARPES,FarrellTrARPES}\begin{align}
			I(\bm{k},\epsilon,t_d)={}&{\rm Im}\int dt_1\int dt_2 \, s(t_1,t_d)s(t_2,t_d)\notag\\
			&\times e^{i\epsilon(t_1-t_2)}{\rm Tr}[G^<(\bm{k},t_1,t_2)]\,,
		\end{align}where $\epsilon$ is the energy of the photoemitted electron measured by the detector and $G^<(\bm{k},t_1,t_2)$ is the lesser Green's function. The spin-resolved photocurrent is naturally extracted by evaluating ${\rm Tr}[s_zG^<(\bm{k},t_1,t_2)]$.

        Figure~\ref{fig3_new} maps the spin-resolved photocurrent across the three distinct parity regimes at $T_{\rm pump}=100$, $T_{\rm probe}=80$, $\mu=0.5$, and $T=0.1$. Under $\omega$--$2\omega$ BLPL driving [Fig.~\ref{fig3_new}(a)] at $S=0.8$, $\phi=\frac{\pi}{2}$, and $\beta=\frac{\pi}{2}$, the spin polarization exhibits purely odd parity, perfectly inverting its sign under $\bm{k}\rightarrow-\bm{k}$. The $\omega$--$2\omega$ BCPL driving [Fig.~\ref{fig3_new}(b)] at $S=0.6$, $\phi=0$, and $\eta_1 = \eta_2=1$, breaks this antisymmetry, producing a mixed-parity photocurrent. Finally, the higher-order $\omega$--$3\omega$ BCLPL driving [Fig.~\ref{fig3_new}(c)] at $S=0.6$, $\eta_1=1$, $\phi=\frac{\pi}{3}$, and $\beta=\frac{\pi}{4}$, enforces an even-parity state, rendering the spin texture invariant upon momentum inversion.
        
			\blue{\textit{Conclusion}}---In summary, we demonstrate that bichromatic $\omega$--$n\omega$ Floquet driving provides a deterministic route to engineer parity-selective spin splittings in a coplanar AFM, bypassing its intrinsic symmetry constraints. Notably, the great flexibility of the driving protocols (BLPL, BCPL, BCLPL) and the availability of multiple tunable parameters enable on-demand control over the spin splitting, including both its parity (even, odd, and mixed) and its wave-type character ($s$, $p$, and $d$). Experimentally, these unconventional spin-splitting states are readily observable, either through distinct transport signatures or via direct visualization using spin-resolved ARPES. By enabling selective access to odd-, even-, and mixed-parity spin-split states within a single AFM, our framework establishes a unified approach to the ultrafast control of unidirectional spin polarization. This dynamic parity tuning provides a versatile platform for systematically exploring parity-dependent transport, unconventional superconductivity, and other emergent nonequilibrium phenomena.
			
			\blue{\textit{Acknowledgments}}---D.\,Z. and Z.\,Y. were supported by the Fundamental and Interdisciplinary Disciplines Breakthrough Plan of the Ministry of Education of China (Grant No. JYB2025XDXM403) and the Guangdong Basic and Applied Basic Research Foundation (Grant No. 2023B1515040023). M.\,Y. was supported by the Department of Energy, Office of Basic Energy Sciences, Division of Materials Sciences and Engineering under Contract No. DE-FG02-08ER46542 for the formal developments, the analytical/numerical work, and the writing of the manuscript. 
			
		}
        
		\bibliography{bib_v2.bib}

\begin{thebibliography}{80}%
\makeatletter
\providecommand \@ifxundefined [1]{%
 \@ifx{#1\undefined}
}%
\providecommand \@ifnum [1]{%
 \ifnum #1\expandafter \@firstoftwo
 \else \expandafter \@secondoftwo
 \fi
}%
\providecommand \@ifx [1]{%
 \ifx #1\expandafter \@firstoftwo
 \else \expandafter \@secondoftwo
 \fi
}%
\providecommand \natexlab [1]{#1}%
\providecommand \enquote  [1]{``#1''}%
\providecommand \bibnamefont  [1]{#1}%
\providecommand \bibfnamefont [1]{#1}%
\providecommand \citenamefont [1]{#1}%
\providecommand \href@noop [0]{\@secondoftwo}%
\providecommand \href [0]{\begingroup \@sanitize@url \@href}%
\providecommand \@href[1]{\@@startlink{#1}\@@href}%
\providecommand \@@href[1]{\endgroup#1\@@endlink}%
\providecommand \@sanitize@url [0]{\catcode `\\12\catcode `\$12\catcode
  `\&12\catcode `\#12\catcode `\^12\catcode `\_12\catcode `\%12\relax}%
\providecommand \@@startlink[1]{}%
\providecommand \@@endlink[0]{}%
\providecommand \url  [0]{\begingroup\@sanitize@url \@url }%
\providecommand \@url [1]{\endgroup\@href {#1}{\urlprefix }}%
\providecommand \urlprefix  [0]{URL }%
\providecommand \Eprint [0]{\href }%
\providecommand \doibase [0]{https://doi.org/}%
\providecommand \selectlanguage [0]{\@gobble}%
\providecommand \bibinfo  [0]{\@secondoftwo}%
\providecommand \bibfield  [0]{\@secondoftwo}%
\providecommand \translation [1]{[#1]}%
\providecommand \BibitemOpen [0]{}%
\providecommand \bibitemStop [0]{}%
\providecommand \bibitemNoStop [0]{.\EOS\space}%
\providecommand \EOS [0]{\spacefactor3000\relax}%
\providecommand \BibitemShut  [1]{\csname bibitem#1\endcsname}%
\let\auto@bib@innerbib\@empty
\bibitem [{\citenamefont {Hayami}\ \emph {et~al.}(2019)\citenamefont {Hayami},
  \citenamefont {Yanagi},\ and\ \citenamefont {Kusunose}}]{Hayami2019AM}%
  \BibitemOpen
  \bibfield  {author} {\bibinfo {author} {\bibfnamefont {S.}~\bibnamefont
  {Hayami}}, \bibinfo {author} {\bibfnamefont {Y.}~\bibnamefont {Yanagi}},\
  and\ \bibinfo {author} {\bibfnamefont {H.}~\bibnamefont {Kusunose}},\
  }\bibfield  {title} {\bibinfo {title} {Momentum-dependent spin splitting by
  collinear antiferromagnetic ordering},\ }\href
  {https://doi.org/10.7566/JPSJ.88.123702} {\bibfield  {journal} {\bibinfo
  {journal} {Journal of the Physical Society of Japan}\ }\textbf {\bibinfo
  {volume} {88}},\ \bibinfo {pages} {123702} (\bibinfo {year}
  {2019})}\BibitemShut {NoStop}%
\bibitem [{\citenamefont {Yuan}\ \emph {et~al.}(2020)\citenamefont {Yuan},
  \citenamefont {Wang}, \citenamefont {Luo}, \citenamefont {Rashba},\ and\
  \citenamefont {Zunger}}]{LDYuan2020}%
  \BibitemOpen
  \bibfield  {author} {\bibinfo {author} {\bibfnamefont {L.-D.}\ \bibnamefont
  {Yuan}}, \bibinfo {author} {\bibfnamefont {Z.}~\bibnamefont {Wang}}, \bibinfo
  {author} {\bibfnamefont {J.-W.}\ \bibnamefont {Luo}}, \bibinfo {author}
  {\bibfnamefont {E.~I.}\ \bibnamefont {Rashba}},\ and\ \bibinfo {author}
  {\bibfnamefont {A.}~\bibnamefont {Zunger}},\ }\bibfield  {title} {\bibinfo
  {title} {Giant momentum-dependent spin splitting in centrosymmetric low-${Z}$
  antiferromagnets},\ }\href {https://doi.org/10.1103/PhysRevB.102.014422}
  {\bibfield  {journal} {\bibinfo  {journal} {Phys. Rev. B}\ }\textbf {\bibinfo
  {volume} {102}},\ \bibinfo {pages} {014422} (\bibinfo {year}
  {2020})}\BibitemShut {NoStop}%
\bibitem [{\citenamefont {Yuan}\ \emph {et~al.}(2021)\citenamefont {Yuan},
  \citenamefont {Wang}, \citenamefont {Luo},\ and\ \citenamefont
  {Zunger}}]{LDYuan2021}%
  \BibitemOpen
  \bibfield  {author} {\bibinfo {author} {\bibfnamefont {L.-D.}\ \bibnamefont
  {Yuan}}, \bibinfo {author} {\bibfnamefont {Z.}~\bibnamefont {Wang}}, \bibinfo
  {author} {\bibfnamefont {J.-W.}\ \bibnamefont {Luo}},\ and\ \bibinfo {author}
  {\bibfnamefont {A.}~\bibnamefont {Zunger}},\ }\bibfield  {title} {\bibinfo
  {title} {Prediction of low-${Z}$ collinear and noncollinear antiferromagnetic
  compounds having momentum-dependent spin splitting even without spin-orbit
  coupling},\ }\href {https://doi.org/10.1103/PhysRevMaterials.5.014409}
  {\bibfield  {journal} {\bibinfo  {journal} {Phys. Rev. Mater.}\ }\textbf
  {\bibinfo {volume} {5}},\ \bibinfo {pages} {014409} (\bibinfo {year}
  {2021})}\BibitemShut {NoStop}%
\bibitem [{\citenamefont {Ma}\ \emph {et~al.}(2021)\citenamefont {Ma},
  \citenamefont {Hu}, \citenamefont {Li}, \citenamefont {Liu}, \citenamefont
  {Yao}, \citenamefont {Jia},\ and\ \citenamefont {Liu}}]{Ma2021AM}%
  \BibitemOpen
  \bibfield  {author} {\bibinfo {author} {\bibfnamefont {H.-Y.}\ \bibnamefont
  {Ma}}, \bibinfo {author} {\bibfnamefont {M.}~\bibnamefont {Hu}}, \bibinfo
  {author} {\bibfnamefont {N.}~\bibnamefont {Li}}, \bibinfo {author}
  {\bibfnamefont {J.}~\bibnamefont {Liu}}, \bibinfo {author} {\bibfnamefont
  {W.}~\bibnamefont {Yao}}, \bibinfo {author} {\bibfnamefont {J.-F.}\
  \bibnamefont {Jia}},\ and\ \bibinfo {author} {\bibfnamefont {J.}~\bibnamefont
  {Liu}},\ }\bibfield  {title} {\bibinfo {title} {Multifunctional
  antiferromagnetic materials with giant piezomagnetism and noncollinear spin
  current},\ }\href {https://doi.org/10.1038/s41467-021-23127-7} {\bibfield
  {journal} {\bibinfo  {journal} {Nature Communications}\ }\textbf {\bibinfo
  {volume} {12}},\ \bibinfo {pages} {2846} (\bibinfo {year}
  {2021})}\BibitemShut {NoStop}%
\bibitem [{\citenamefont {\ifmmode~\check{S}\else \v{S}\fi{}mejkal}\ \emph
  {et~al.}(2022{\natexlab{a}})\citenamefont {\ifmmode~\check{S}\else
  \v{S}\fi{}mejkal}, \citenamefont {Sinova},\ and\ \citenamefont
  {Jungwirth}}]{libor2022AMa}%
  \BibitemOpen
  \bibfield  {author} {\bibinfo {author} {\bibfnamefont {L.}~\bibnamefont
  {\ifmmode~\check{S}\else \v{S}\fi{}mejkal}}, \bibinfo {author} {\bibfnamefont
  {J.}~\bibnamefont {Sinova}},\ and\ \bibinfo {author} {\bibfnamefont
  {T.}~\bibnamefont {Jungwirth}},\ }\bibfield  {title} {\bibinfo {title}
  {Beyond conventional ferromagnetism and antiferromagnetism: A phase with
  nonrelativistic spin and crystal rotation symmetry},\ }\href
  {https://doi.org/10.1103/PhysRevX.12.031042} {\bibfield  {journal} {\bibinfo
  {journal} {Phys. Rev. X}\ }\textbf {\bibinfo {volume} {12}},\ \bibinfo
  {pages} {031042} (\bibinfo {year} {2022}{\natexlab{a}})}\BibitemShut
  {NoStop}%
\bibitem [{\citenamefont {\ifmmode~\check{S}\else \v{S}\fi{}mejkal}\ \emph
  {et~al.}(2022{\natexlab{b}})\citenamefont {\ifmmode~\check{S}\else
  \v{S}\fi{}mejkal}, \citenamefont {Sinova},\ and\ \citenamefont
  {Jungwirth}}]{Libor2022AMb}%
  \BibitemOpen
  \bibfield  {author} {\bibinfo {author} {\bibfnamefont {L.}~\bibnamefont
  {\ifmmode~\check{S}\else \v{S}\fi{}mejkal}}, \bibinfo {author} {\bibfnamefont
  {J.}~\bibnamefont {Sinova}},\ and\ \bibinfo {author} {\bibfnamefont
  {T.}~\bibnamefont {Jungwirth}},\ }\bibfield  {title} {\bibinfo {title}
  {Emerging research landscape of altermagnetism},\ }\href
  {https://doi.org/10.1103/PhysRevX.12.040501} {\bibfield  {journal} {\bibinfo
  {journal} {Phys. Rev. X}\ }\textbf {\bibinfo {volume} {12}},\ \bibinfo
  {pages} {040501} (\bibinfo {year} {2022}{\natexlab{b}})}\BibitemShut
  {NoStop}%
\bibitem [{\citenamefont {\ifmmode~\check{S}\else \v{S}\fi{}mejkal}\ \emph
  {et~al.}(2022{\natexlab{c}})\citenamefont {\ifmmode~\check{S}\else
  \v{S}\fi{}mejkal}, \citenamefont {Hellenes}, \citenamefont
  {Gonz\'alez-Hern\'andez}, \citenamefont {Sinova},\ and\ \citenamefont
  {Jungwirth}}]{Libor2022AMc}%
  \BibitemOpen
  \bibfield  {author} {\bibinfo {author} {\bibfnamefont {L.}~\bibnamefont
  {\ifmmode~\check{S}\else \v{S}\fi{}mejkal}}, \bibinfo {author} {\bibfnamefont
  {A.~B.}\ \bibnamefont {Hellenes}}, \bibinfo {author} {\bibfnamefont
  {R.}~\bibnamefont {Gonz\'alez-Hern\'andez}}, \bibinfo {author} {\bibfnamefont
  {J.}~\bibnamefont {Sinova}},\ and\ \bibinfo {author} {\bibfnamefont
  {T.}~\bibnamefont {Jungwirth}},\ }\bibfield  {title} {\bibinfo {title} {Giant
  and tunneling magnetoresistance in unconventional collinear antiferromagnets
  with nonrelativistic spin-momentum coupling},\ }\href
  {https://doi.org/10.1103/PhysRevX.12.011028} {\bibfield  {journal} {\bibinfo
  {journal} {Phys. Rev. X}\ }\textbf {\bibinfo {volume} {12}},\ \bibinfo
  {pages} {011028} (\bibinfo {year} {2022}{\natexlab{c}})}\BibitemShut
  {NoStop}%
\bibitem [{\citenamefont {Song}\ \emph {et~al.}(2025)\citenamefont {Song},
  \citenamefont {Qi}, \citenamefont {Fang}, \citenamefont {Fang},\ and\
  \citenamefont {Weng}}]{song2025AM}%
  \BibitemOpen
  \bibfield  {author} {\bibinfo {author} {\bibfnamefont {Z.}~\bibnamefont
  {Song}}, \bibinfo {author} {\bibfnamefont {Z.}~\bibnamefont {Qi}}, \bibinfo
  {author} {\bibfnamefont {C.}~\bibnamefont {Fang}}, \bibinfo {author}
  {\bibfnamefont {Z.}~\bibnamefont {Fang}},\ and\ \bibinfo {author}
  {\bibfnamefont {H.}~\bibnamefont {Weng}},\ }\bibfield  {title} {\bibinfo
  {title} {A unified symmetry classification of magnetic orders via spin space
  groups: Prediction of coplanar even-wave phases},\ }\href
  {https://arxiv.org/abs/2512.08901} {\  (\bibinfo {year} {2025})},\ \Eprint
  {https://arxiv.org/abs/2512.08901} {arXiv:2512.08901} \BibitemShut {NoStop}%
\bibitem [{\citenamefont {Hellenes}\ \emph {et~al.}(2024)\citenamefont
  {Hellenes}, \citenamefont {Jungwirth}, \citenamefont {Jaeschke-Ubiergo},
  \citenamefont {Chakraborty}, \citenamefont {Sinova},\ and\ \citenamefont
  {Šmejkal}}]{Birk2023}%
  \BibitemOpen
  \bibfield  {author} {\bibinfo {author} {\bibfnamefont {A.~B.}\ \bibnamefont
  {Hellenes}}, \bibinfo {author} {\bibfnamefont {T.}~\bibnamefont {Jungwirth}},
  \bibinfo {author} {\bibfnamefont {R.}~\bibnamefont {Jaeschke-Ubiergo}},
  \bibinfo {author} {\bibfnamefont {A.}~\bibnamefont {Chakraborty}}, \bibinfo
  {author} {\bibfnamefont {J.}~\bibnamefont {Sinova}},\ and\ \bibinfo {author}
  {\bibfnamefont {L.}~\bibnamefont {Šmejkal}},\ }\bibfield  {title} {\bibinfo
  {title} {P-wave magnets},\ }\href {https://arxiv.org/abs/2309.01607} {\
  (\bibinfo {year} {2024})},\ \Eprint {https://arxiv.org/abs/2309.01607}
  {arXiv:2309.01607} \BibitemShut {NoStop}%
\bibitem [{\citenamefont {Lin}\ and\ \citenamefont {Vila}(2026)}]{Lin2025OAM}%
  \BibitemOpen
  \bibfield  {author} {\bibinfo {author} {\bibfnamefont {Y.-P.}\ \bibnamefont
  {Lin}}\ and\ \bibinfo {author} {\bibfnamefont {M.}~\bibnamefont {Vila}},\
  }\bibfield  {title} {\bibinfo {title} {Odd-parity altermagnetism through
  sublattice currents: From haldane-hubbard model to general bipartite
  lattices},\ }\href {https://arxiv.org/abs/2503.09602} {\  (\bibinfo {year}
  {2026})},\ \Eprint {https://arxiv.org/abs/2503.09602} {arXiv:2503.09602}
  \BibitemShut {NoStop}%
\bibitem [{\citenamefont {Yu}\ \emph {et~al.}(2025)\citenamefont {Yu},
  \citenamefont {Lyngby}, \citenamefont {Shishidou}, \citenamefont {Roig},
  \citenamefont {Kreisel}, \citenamefont {Weinert}, \citenamefont {Andersen},\
  and\ \citenamefont {Agterberg}}]{Yu2025pwave}%
  \BibitemOpen
  \bibfield  {author} {\bibinfo {author} {\bibfnamefont {Y.}~\bibnamefont
  {Yu}}, \bibinfo {author} {\bibfnamefont {M.~B.}\ \bibnamefont {Lyngby}},
  \bibinfo {author} {\bibfnamefont {T.}~\bibnamefont {Shishidou}}, \bibinfo
  {author} {\bibfnamefont {M.}~\bibnamefont {Roig}}, \bibinfo {author}
  {\bibfnamefont {A.}~\bibnamefont {Kreisel}}, \bibinfo {author} {\bibfnamefont
  {M.}~\bibnamefont {Weinert}}, \bibinfo {author} {\bibfnamefont {B.~M.}\
  \bibnamefont {Andersen}},\ and\ \bibinfo {author} {\bibfnamefont {D.~F.}\
  \bibnamefont {Agterberg}},\ }\bibfield  {title} {\bibinfo {title} {Odd-parity
  magnetism driven by antiferromagnetic exchange},\ }\href
  {https://doi.org/10.1103/zk69-k6b2} {\bibfield  {journal} {\bibinfo
  {journal} {Phys. Rev. Lett.}\ }\textbf {\bibinfo {volume} {135}},\ \bibinfo
  {pages} {046701} (\bibinfo {year} {2025})}\BibitemShut {NoStop}%
\bibitem [{\citenamefont {Zhuang}\ \emph
  {et~al.}(2026{\natexlab{a}})\citenamefont {Zhuang}, \citenamefont {Zhu},
  \citenamefont {Liu}, \citenamefont {Wu},\ and\ \citenamefont
  {Yan}}]{zhuang2025AM9}%
  \BibitemOpen
  \bibfield  {author} {\bibinfo {author} {\bibfnamefont {Z.-Y.}\ \bibnamefont
  {Zhuang}}, \bibinfo {author} {\bibfnamefont {D.}~\bibnamefont {Zhu}},
  \bibinfo {author} {\bibfnamefont {D.}~\bibnamefont {Liu}}, \bibinfo {author}
  {\bibfnamefont {Z.}~\bibnamefont {Wu}},\ and\ \bibinfo {author}
  {\bibfnamefont {Z.}~\bibnamefont {Yan}},\ }\bibfield  {title} {\bibinfo
  {title} {Odd-parity altermagnetism originated from orbital orders},\ }\href
  {https://arxiv.org/abs/2508.18361} {\  (\bibinfo {year}
  {2026}{\natexlab{a}})},\ \Eprint {https://arxiv.org/abs/2508.18361}
  {arXiv:2508.18361} \BibitemShut {NoStop}%
\bibitem [{\citenamefont {Luo}\ \emph {et~al.}(2026{\natexlab{a}})\citenamefont
  {Luo}, \citenamefont {Hu}, \citenamefont {Hu},\ and\ \citenamefont
  {Law}}]{Luo2026UAFM1}%
  \BibitemOpen
  \bibfield  {author} {\bibinfo {author} {\bibfnamefont {X.-J.}\ \bibnamefont
  {Luo}}, \bibinfo {author} {\bibfnamefont {J.-X.}\ \bibnamefont {Hu}},
  \bibinfo {author} {\bibfnamefont {M.}~\bibnamefont {Hu}},\ and\ \bibinfo
  {author} {\bibfnamefont {K.~T.}\ \bibnamefont {Law}},\ }\bibfield  {title}
  {\bibinfo {title} {Spin group symmetry criteria for unconventional
  magnetism},\ }\href {https://arxiv.org/abs/2603.07643} {\  (\bibinfo {year}
  {2026}{\natexlab{a}})},\ \Eprint {https://arxiv.org/abs/2603.07643}
  {arXiv:2603.07643} \BibitemShut {NoStop}%
\bibitem [{\citenamefont {Luo}\ \emph {et~al.}(2026{\natexlab{b}})\citenamefont
  {Luo}, \citenamefont {Li}, \citenamefont {Xiao}, \citenamefont {Shao},
  \citenamefont {Li}, \citenamefont {Tian},\ and\ \citenamefont
  {Yao}}]{Luo2026UAFM2}%
  \BibitemOpen
  \bibfield  {author} {\bibinfo {author} {\bibfnamefont {X.-J.}\ \bibnamefont
  {Luo}}, \bibinfo {author} {\bibfnamefont {D.}~\bibnamefont {Li}}, \bibinfo
  {author} {\bibfnamefont {R.-C.}\ \bibnamefont {Xiao}}, \bibinfo {author}
  {\bibfnamefont {D.-F.}\ \bibnamefont {Shao}}, \bibinfo {author}
  {\bibfnamefont {L.}~\bibnamefont {Li}}, \bibinfo {author} {\bibfnamefont
  {M.}~\bibnamefont {Tian}},\ and\ \bibinfo {author} {\bibfnamefont
  {Y.}~\bibnamefont {Yao}},\ }\bibfield  {title} {\bibinfo {title}
  {Unconventional magnetism: Symmetry classification, hybrid-parity and
  unconstrained-parity classes},\ }\href {https://arxiv.org/abs/2605.21336} {\
  (\bibinfo {year} {2026}{\natexlab{b}})},\ \Eprint
  {https://arxiv.org/abs/2605.21336} {arXiv:2605.21336} \BibitemShut {NoStop}%
\bibitem [{\citenamefont {Zeng}\ \emph {et~al.}(2026)\citenamefont {Zeng},
  \citenamefont {Qin}, \citenamefont {Qin}, \citenamefont {Feng}, \citenamefont
  {Wu}, \citenamefont {Xu},\ and\ \citenamefont {Wang}}]{Zeng2026OddAM}%
  \BibitemOpen
  \bibfield  {author} {\bibinfo {author} {\bibfnamefont {M.}~\bibnamefont
  {Zeng}}, \bibinfo {author} {\bibfnamefont {Z.}~\bibnamefont {Qin}}, \bibinfo
  {author} {\bibfnamefont {L.}~\bibnamefont {Qin}}, \bibinfo {author}
  {\bibfnamefont {S.}~\bibnamefont {Feng}}, \bibinfo {author} {\bibfnamefont
  {L.}~\bibnamefont {Wu}}, \bibinfo {author} {\bibfnamefont {D.-H.}\
  \bibnamefont {Xu}},\ and\ \bibinfo {author} {\bibfnamefont {R.}~\bibnamefont
  {Wang}},\ }\bibfield  {title} {\bibinfo {title} {Odd-parity altermagnetism: A
  spin group study},\ }\href {https://doi.org/10.1103/7kmk-yl2t} {\bibfield
  {journal} {\bibinfo  {journal} {Phys. Rev. B}\ }\textbf {\bibinfo {volume}
  {113}},\ \bibinfo {pages} {L220412} (\bibinfo {year} {2026})}\BibitemShut
  {NoStop}%
\bibitem [{\citenamefont {Shao}\ \emph {et~al.}(2021)\citenamefont {Shao},
  \citenamefont {Zhang}, \citenamefont {Li}, \citenamefont {Eom},\ and\
  \citenamefont {Tsymbal}}]{Shao2021NC}%
  \BibitemOpen
  \bibfield  {author} {\bibinfo {author} {\bibfnamefont {D.-F.}\ \bibnamefont
  {Shao}}, \bibinfo {author} {\bibfnamefont {S.-H.}\ \bibnamefont {Zhang}},
  \bibinfo {author} {\bibfnamefont {M.}~\bibnamefont {Li}}, \bibinfo {author}
  {\bibfnamefont {C.-B.}\ \bibnamefont {Eom}},\ and\ \bibinfo {author}
  {\bibfnamefont {E.~Y.}\ \bibnamefont {Tsymbal}},\ }\bibfield  {title}
  {\bibinfo {title} {Spin-neutral currents for spintronics},\ }\href
  {https://doi.org/10.1038/s41467-021-26915-3} {\bibfield  {journal} {\bibinfo
  {journal} {Nature Communications}\ }\textbf {\bibinfo {volume} {12}},\
  \bibinfo {pages} {7061} (\bibinfo {year} {2021})}\BibitemShut {NoStop}%
\bibitem [{\citenamefont {Ouassou}\ \emph {et~al.}(2023)\citenamefont
  {Ouassou}, \citenamefont {Brataas},\ and\ \citenamefont
  {Linder}}]{Ouassou2023AM}%
  \BibitemOpen
  \bibfield  {author} {\bibinfo {author} {\bibfnamefont {J.~A.}\ \bibnamefont
  {Ouassou}}, \bibinfo {author} {\bibfnamefont {A.}~\bibnamefont {Brataas}},\
  and\ \bibinfo {author} {\bibfnamefont {J.}~\bibnamefont {Linder}},\
  }\bibfield  {title} {\bibinfo {title} {dc {J}osephson effect in
  altermagnets},\ }\href {https://doi.org/10.1103/PhysRevLett.131.076003}
  {\bibfield  {journal} {\bibinfo  {journal} {Phys. Rev. Lett.}\ }\textbf
  {\bibinfo {volume} {131}},\ \bibinfo {pages} {076003} (\bibinfo {year}
  {2023})}\BibitemShut {NoStop}%
\bibitem [{\citenamefont {Lu}\ \emph {et~al.}(2024)\citenamefont {Lu},
  \citenamefont {Maeda}, \citenamefont {Ito}, \citenamefont {Yada},\ and\
  \citenamefont {Tanaka}}]{Lu2024AM}%
  \BibitemOpen
  \bibfield  {author} {\bibinfo {author} {\bibfnamefont {B.}~\bibnamefont
  {Lu}}, \bibinfo {author} {\bibfnamefont {K.}~\bibnamefont {Maeda}}, \bibinfo
  {author} {\bibfnamefont {H.}~\bibnamefont {Ito}}, \bibinfo {author}
  {\bibfnamefont {K.}~\bibnamefont {Yada}},\ and\ \bibinfo {author}
  {\bibfnamefont {Y.}~\bibnamefont {Tanaka}},\ }\bibfield  {title} {\bibinfo
  {title} {$\ensuremath{\varphi}$ {J}osephson junction induced by
  altermagnetism},\ }\href {https://doi.org/10.1103/PhysRevLett.133.226002}
  {\bibfield  {journal} {\bibinfo  {journal} {Phys. Rev. Lett.}\ }\textbf
  {\bibinfo {volume} {133}},\ \bibinfo {pages} {226002} (\bibinfo {year}
  {2024})}\BibitemShut {NoStop}%
\bibitem [{\citenamefont {Lin}\ \emph {et~al.}(2025)\citenamefont {Lin},
  \citenamefont {Zhang}, \citenamefont {Lu},\ and\ \citenamefont
  {Xie}}]{Lin2024AM12}%
  \BibitemOpen
  \bibfield  {author} {\bibinfo {author} {\bibfnamefont {H.-J.}\ \bibnamefont
  {Lin}}, \bibinfo {author} {\bibfnamefont {S.-B.}\ \bibnamefont {Zhang}},
  \bibinfo {author} {\bibfnamefont {H.-Z.}\ \bibnamefont {Lu}},\ and\ \bibinfo
  {author} {\bibfnamefont {X.~C.}\ \bibnamefont {Xie}},\ }\bibfield  {title}
  {\bibinfo {title} {Coulomb drag in altermagnets},\ }\href
  {https://doi.org/10.1103/PhysRevLett.134.136301} {\bibfield  {journal}
  {\bibinfo  {journal} {Phys. Rev. Lett.}\ }\textbf {\bibinfo {volume} {134}},\
  \bibinfo {pages} {136301} (\bibinfo {year} {2025})}\BibitemShut {NoStop}%
\bibitem [{\citenamefont {Zhuang}\ \emph
  {et~al.}(2026{\natexlab{b}})\citenamefont {Zhuang}, \citenamefont {Zhu},
  \citenamefont {Wu},\ and\ \citenamefont {Yan}}]{Zhuang2025SNL}%
  \BibitemOpen
  \bibfield  {author} {\bibinfo {author} {\bibfnamefont {Z.-Y.}\ \bibnamefont
  {Zhuang}}, \bibinfo {author} {\bibfnamefont {D.}~\bibnamefont {Zhu}},
  \bibinfo {author} {\bibfnamefont {Z.}~\bibnamefont {Wu}},\ and\ \bibinfo
  {author} {\bibfnamefont {Z.}~\bibnamefont {Yan}},\ }\bibfield  {title}
  {\bibinfo {title} {Cartesian nodal lines and magnetic kramers {W}eyl nodes in
  spin-split antiferromagnets},\ }\href
  {https://doi.org/10.1016/j.newton.2026.100403} {\bibfield  {journal}
  {\bibinfo  {journal} {Newton}\ }\textbf {\bibinfo {volume} {2}},\ \bibinfo
  {pages} {100403} (\bibinfo {year} {2026}{\natexlab{b}})}\BibitemShut
  {NoStop}%
\bibitem [{\citenamefont {Chakraborty}\ and\ \citenamefont
  {Black-Schaffer}(2025{\natexlab{a}})}]{Chakraborty2025AM}%
  \BibitemOpen
  \bibfield  {author} {\bibinfo {author} {\bibfnamefont {D.}~\bibnamefont
  {Chakraborty}}\ and\ \bibinfo {author} {\bibfnamefont {A.~M.}\ \bibnamefont
  {Black-Schaffer}},\ }\bibfield  {title} {\bibinfo {title} {Perfect
  superconducting diode effect in altermagnets},\ }\href
  {https://doi.org/10.1103/cv8s-tk4c} {\bibfield  {journal} {\bibinfo
  {journal} {Phys. Rev. Lett.}\ }\textbf {\bibinfo {volume} {135}},\ \bibinfo
  {pages} {026001} (\bibinfo {year} {2025}{\natexlab{a}})}\BibitemShut
  {NoStop}%
\bibitem [{\citenamefont {Brekke}\ \emph {et~al.}(2024)\citenamefont {Brekke},
  \citenamefont {Sukhachov}, \citenamefont {Giil}, \citenamefont {Brataas},\
  and\ \citenamefont {Linder}}]{Brekke2024pwave}%
  \BibitemOpen
  \bibfield  {author} {\bibinfo {author} {\bibfnamefont {B.}~\bibnamefont
  {Brekke}}, \bibinfo {author} {\bibfnamefont {P.}~\bibnamefont {Sukhachov}},
  \bibinfo {author} {\bibfnamefont {H.~G.}\ \bibnamefont {Giil}}, \bibinfo
  {author} {\bibfnamefont {A.}~\bibnamefont {Brataas}},\ and\ \bibinfo {author}
  {\bibfnamefont {J.}~\bibnamefont {Linder}},\ }\bibfield  {title} {\bibinfo
  {title} {Minimal models and transport properties of unconventional $p$-wave
  magnets},\ }\href {https://doi.org/10.1103/PhysRevLett.133.236703} {\bibfield
   {journal} {\bibinfo  {journal} {Phys. Rev. Lett.}\ }\textbf {\bibinfo
  {volume} {133}},\ \bibinfo {pages} {236703} (\bibinfo {year}
  {2024})}\BibitemShut {NoStop}%
\bibitem [{\citenamefont {Hedayati}\ and\ \citenamefont
  {Salehi}(2025)}]{Hedayati2025pwave}%
  \BibitemOpen
  \bibfield  {author} {\bibinfo {author} {\bibfnamefont {A.~A.}\ \bibnamefont
  {Hedayati}}\ and\ \bibinfo {author} {\bibfnamefont {M.}~\bibnamefont
  {Salehi}},\ }\bibfield  {title} {\bibinfo {title} {Transverse spin current at
  normal-metal /$p$-wave magnet junctions},\ }\href
  {https://doi.org/10.1103/PhysRevB.111.035404} {\bibfield  {journal} {\bibinfo
   {journal} {Phys. Rev. B}\ }\textbf {\bibinfo {volume} {111}},\ \bibinfo
  {pages} {035404} (\bibinfo {year} {2025})}\BibitemShut {NoStop}%
\bibitem [{\citenamefont {Ezawa}(2025)}]{Ezawa2025pwave}%
  \BibitemOpen
  \bibfield  {author} {\bibinfo {author} {\bibfnamefont {M.}~\bibnamefont
  {Ezawa}},\ }\bibfield  {title} {\bibinfo {title} {Out-of-plane edelstein
  effects: Electric field induced magnetization in $p$-wave magnets},\ }\href
  {https://doi.org/10.1103/PhysRevB.111.L161301} {\bibfield  {journal}
  {\bibinfo  {journal} {Phys. Rev. B}\ }\textbf {\bibinfo {volume} {111}},\
  \bibinfo {pages} {L161301} (\bibinfo {year} {2025})}\BibitemShut {NoStop}%
\bibitem [{\citenamefont {Zhuang}\ \emph
  {et~al.}(2026{\natexlab{c}})\citenamefont {Zhuang}, \citenamefont {Hu},
  \citenamefont {Zhang}, \citenamefont {Hu},\ and\ \citenamefont
  {Yan}}]{zhuang2026Mixed}%
  \BibitemOpen
  \bibfield  {author} {\bibinfo {author} {\bibfnamefont {Z.-Y.}\ \bibnamefont
  {Zhuang}}, \bibinfo {author} {\bibfnamefont {J.-X.}\ \bibnamefont {Hu}},
  \bibinfo {author} {\bibfnamefont {S.-B.}\ \bibnamefont {Zhang}}, \bibinfo
  {author} {\bibfnamefont {L.-H.}\ \bibnamefont {Hu}},\ and\ \bibinfo {author}
  {\bibfnamefont {Z.}~\bibnamefont {Yan}},\ }\bibfield  {title} {\bibinfo
  {title} {Mixed-parity altermagnetism in collinear spin-orbital magnets},\
  }\href {https://arxiv.org/abs/2605.05205} {\  (\bibinfo {year}
  {2026}{\natexlab{c}})},\ \Eprint {https://arxiv.org/abs/2605.05205}
  {arXiv:2605.05205} \BibitemShut {NoStop}%
\bibitem [{\citenamefont {Zhu}\ \emph {et~al.}(2023)\citenamefont {Zhu},
  \citenamefont {Zhuang}, \citenamefont {Wu},\ and\ \citenamefont
  {Yan}}]{Zhu2023TSC}%
  \BibitemOpen
  \bibfield  {author} {\bibinfo {author} {\bibfnamefont {D.}~\bibnamefont
  {Zhu}}, \bibinfo {author} {\bibfnamefont {Z.-Y.}\ \bibnamefont {Zhuang}},
  \bibinfo {author} {\bibfnamefont {Z.}~\bibnamefont {Wu}},\ and\ \bibinfo
  {author} {\bibfnamefont {Z.}~\bibnamefont {Yan}},\ }\bibfield  {title}
  {\bibinfo {title} {Topological superconductivity in two-dimensional
  altermagnetic metals},\ }\href {https://doi.org/10.1103/PhysRevB.108.184505}
  {\bibfield  {journal} {\bibinfo  {journal} {Phys. Rev. B}\ }\textbf {\bibinfo
  {volume} {108}},\ \bibinfo {pages} {184505} (\bibinfo {year}
  {2023})}\BibitemShut {NoStop}%
\bibitem [{\citenamefont {Brekke}\ \emph {et~al.}(2023)\citenamefont {Brekke},
  \citenamefont {Brataas},\ and\ \citenamefont {Sudb\o{}}}]{Brekke2023AM}%
  \BibitemOpen
  \bibfield  {author} {\bibinfo {author} {\bibfnamefont {B.}~\bibnamefont
  {Brekke}}, \bibinfo {author} {\bibfnamefont {A.}~\bibnamefont {Brataas}},\
  and\ \bibinfo {author} {\bibfnamefont {A.}~\bibnamefont {Sudb\o{}}},\
  }\bibfield  {title} {\bibinfo {title} {{Two-dimensional altermagnets:
  Superconductivity in a minimal microscopic model}},\ }\href
  {https://doi.org/10.1103/PhysRevB.108.224421} {\bibfield  {journal} {\bibinfo
   {journal} {Phys. Rev. B}\ }\textbf {\bibinfo {volume} {108}},\ \bibinfo
  {pages} {224421} (\bibinfo {year} {2023})}\BibitemShut {NoStop}%
\bibitem [{\citenamefont {Zhang}\ \emph {et~al.}(2024)\citenamefont {Zhang},
  \citenamefont {Hu},\ and\ \citenamefont {Neupert}}]{Zhang2024AM}%
  \BibitemOpen
  \bibfield  {author} {\bibinfo {author} {\bibfnamefont {S.-B.}\ \bibnamefont
  {Zhang}}, \bibinfo {author} {\bibfnamefont {L.-H.}\ \bibnamefont {Hu}},\ and\
  \bibinfo {author} {\bibfnamefont {T.}~\bibnamefont {Neupert}},\ }\bibfield
  {title} {\bibinfo {title} {{Finite-momentum Cooper pairing in proximitized
  altermagnets}},\ }\href {https://doi.org/10.1038/s41467-024-45951-3}
  {\bibfield  {journal} {\bibinfo  {journal} {Nature Communications}\ }\textbf
  {\bibinfo {volume} {15}},\ \bibinfo {pages} {1801} (\bibinfo {year}
  {2024})}\BibitemShut {NoStop}%
\bibitem [{\citenamefont {Chakraborty}\ and\ \citenamefont
  {Black-Schaffer}(2024)}]{Chakraborty2024AM}%
  \BibitemOpen
  \bibfield  {author} {\bibinfo {author} {\bibfnamefont {D.}~\bibnamefont
  {Chakraborty}}\ and\ \bibinfo {author} {\bibfnamefont {A.~M.}\ \bibnamefont
  {Black-Schaffer}},\ }\bibfield  {title} {\bibinfo {title} {Zero-field
  finite-momentum and field-induced superconductivity in altermagnets},\ }\href
  {https://doi.org/10.1103/PhysRevB.110.L060508} {\bibfield  {journal}
  {\bibinfo  {journal} {Phys. Rev. B}\ }\textbf {\bibinfo {volume} {110}},\
  \bibinfo {pages} {L060508} (\bibinfo {year} {2024})}\BibitemShut {NoStop}%
\bibitem [{\citenamefont {Chakraborty}\ and\ \citenamefont
  {Black-Schaffer}(2025{\natexlab{b}})}]{Chakraborty2025AM2}%
  \BibitemOpen
  \bibfield  {author} {\bibinfo {author} {\bibfnamefont {D.}~\bibnamefont
  {Chakraborty}}\ and\ \bibinfo {author} {\bibfnamefont {A.~M.}\ \bibnamefont
  {Black-Schaffer}},\ }\bibfield  {title} {\bibinfo {title} {Constraints on
  superconducting pairing in altermagnets},\ }\href
  {https://doi.org/10.1103/zylh-rqxl} {\bibfield  {journal} {\bibinfo
  {journal} {Phys. Rev. B}\ }\textbf {\bibinfo {volume} {112}},\ \bibinfo
  {pages} {014516} (\bibinfo {year} {2025}{\natexlab{b}})}\BibitemShut
  {NoStop}%
\bibitem [{\citenamefont {Sun}\ \emph {et~al.}(2025)\citenamefont {Sun},
  \citenamefont {Feng}, \citenamefont {Xie}, \citenamefont {Zhou},
  \citenamefont {Hu},\ and\ \citenamefont {Law}}]{Sun2025pwaveSC}%
  \BibitemOpen
  \bibfield  {author} {\bibinfo {author} {\bibfnamefont {Z.-T.}\ \bibnamefont
  {Sun}}, \bibinfo {author} {\bibfnamefont {X.}~\bibnamefont {Feng}}, \bibinfo
  {author} {\bibfnamefont {Y.-M.}\ \bibnamefont {Xie}}, \bibinfo {author}
  {\bibfnamefont {B.~T.}\ \bibnamefont {Zhou}}, \bibinfo {author}
  {\bibfnamefont {J.-X.}\ \bibnamefont {Hu}},\ and\ \bibinfo {author}
  {\bibfnamefont {K.~T.}\ \bibnamefont {Law}},\ }\bibfield  {title} {\bibinfo
  {title} {Pseudo-{I}sing superconductivity induced by $p$-wave magnetism},\
  }\href {https://doi.org/10.1103/cm1l-1rsh} {\bibfield  {journal} {\bibinfo
  {journal} {Phys. Rev. B}\ }\textbf {\bibinfo {volume} {112}},\ \bibinfo
  {pages} {214504} (\bibinfo {year} {2025})}\BibitemShut {NoStop}%
\bibitem [{\citenamefont {Kim}\ \emph {et~al.}(2026)\citenamefont {Kim},
  \citenamefont {Sim},\ and\ \citenamefont {Park}}]{Kim2026pwaveSC}%
  \BibitemOpen
  \bibfield  {author} {\bibinfo {author} {\bibfnamefont {K.-M.}\ \bibnamefont
  {Kim}}, \bibinfo {author} {\bibfnamefont {G.}~\bibnamefont {Sim}},\ and\
  \bibinfo {author} {\bibfnamefont {M.~J.}\ \bibnamefont {Park}},\ }\bibfield
  {title} {\bibinfo {title} {Topological ising superconductivity in
  two-dimensional p-wave magnet},\ }\href {https://arxiv.org/abs/2605.01686} {\
   (\bibinfo {year} {2026})},\ \Eprint {https://arxiv.org/abs/2605.01686}
  {arXiv:2605.01686} \BibitemShut {NoStop}%
\bibitem [{\citenamefont {Khodas}\ \emph {et~al.}(2026)\citenamefont {Khodas},
  \citenamefont {Šmejkal},\ and\ \citenamefont {Mazin}}]{Khodas2026pwaveSC}%
  \BibitemOpen
  \bibfield  {author} {\bibinfo {author} {\bibfnamefont {M.}~\bibnamefont
  {Khodas}}, \bibinfo {author} {\bibfnamefont {L.}~\bibnamefont {Šmejkal}},\
  and\ \bibinfo {author} {\bibfnamefont {I.~I.}\ \bibnamefont {Mazin}},\
  }\bibfield  {title} {\bibinfo {title} {Nonrelativistic-ising
  superconductivity in p-wave magnets},\ }\href
  {https://arxiv.org/abs/2601.19829} {\  (\bibinfo {year} {2026})},\ \Eprint
  {https://arxiv.org/abs/2601.19829} {arXiv:2601.19829} \BibitemShut {NoStop}%
\bibitem [{\citenamefont {Lee}\ \emph {et~al.}(2024)\citenamefont {Lee},
  \citenamefont {Qian},\ and\ \citenamefont {Yang}}]{Lee2024dwaveNAFM}%
  \BibitemOpen
  \bibfield  {author} {\bibinfo {author} {\bibfnamefont {S.~H.}\ \bibnamefont
  {Lee}}, \bibinfo {author} {\bibfnamefont {Y.}~\bibnamefont {Qian}},\ and\
  \bibinfo {author} {\bibfnamefont {B.-J.}\ \bibnamefont {Yang}},\ }\bibfield
  {title} {\bibinfo {title} {Fermi surface spin texture and topological
  superconductivity in spin-orbit free noncollinear antiferromagnets},\ }\href
  {https://doi.org/10.1103/PhysRevLett.132.196602} {\bibfield  {journal}
  {\bibinfo  {journal} {Phys. Rev. Lett.}\ }\textbf {\bibinfo {volume} {132}},\
  \bibinfo {pages} {196602} (\bibinfo {year} {2024})}\BibitemShut {NoStop}%
\bibitem [{\citenamefont {Zhang}\ and\ \citenamefont
  {Hu}(2025)}]{Zhang2025AMSC}%
  \BibitemOpen
  \bibfield  {author} {\bibinfo {author} {\bibfnamefont {S.-B.}\ \bibnamefont
  {Zhang}}\ and\ \bibinfo {author} {\bibfnamefont {L.-H.}\ \bibnamefont {Hu}},\
  }\bibfield  {title} {\bibinfo {title} {Finite-momentum mixed singlet-triplet
  pairing in chiral antiferromagnets induced by even-parity spin texture},\
  }\href {https://doi.org/10.1103/8cwk-nynj} {\bibfield  {journal} {\bibinfo
  {journal} {Phys. Rev. B}\ }\textbf {\bibinfo {volume} {112}},\ \bibinfo
  {pages} {L100501} (\bibinfo {year} {2025})}\BibitemShut {NoStop}%
\bibitem [{\citenamefont {Zhang}\ \emph {et~al.}(2026)\citenamefont {Zhang},
  \citenamefont {Hu}, \citenamefont {Niu},\ and\ \citenamefont
  {Zhang}}]{zhang2026CAFM}%
  \BibitemOpen
  \bibfield  {author} {\bibinfo {author} {\bibfnamefont {S.-B.}\ \bibnamefont
  {Zhang}}, \bibinfo {author} {\bibfnamefont {L.-H.}\ \bibnamefont {Hu}},
  \bibinfo {author} {\bibfnamefont {Q.}~\bibnamefont {Niu}},\ and\ \bibinfo
  {author} {\bibfnamefont {Z.}~\bibnamefont {Zhang}},\ }\bibfield  {title}
  {\bibinfo {title} {Spin-valley locking and pure spin-triplet
  superconductivity in noncollinear antiferromagnets proximitized to
  conventional superconductors},\ }\href
  {https://doi.org/10.1016/j.newton.2025.100379} {\bibfield  {journal}
  {\bibinfo  {journal} {Newton}\ }\textbf {\bibinfo {volume} {2}},\ \bibinfo
  {pages} {100379} (\bibinfo {year} {2026})}\BibitemShut {NoStop}%
\bibitem [{\citenamefont {Baltz}\ \emph {et~al.}(2018)\citenamefont {Baltz},
  \citenamefont {Manchon}, \citenamefont {Tsoi}, \citenamefont {Moriyama},
  \citenamefont {Ono},\ and\ \citenamefont
  {Tserkovnyak}}]{RevModPhys.90.015005}%
  \BibitemOpen
  \bibfield  {author} {\bibinfo {author} {\bibfnamefont {V.}~\bibnamefont
  {Baltz}}, \bibinfo {author} {\bibfnamefont {A.}~\bibnamefont {Manchon}},
  \bibinfo {author} {\bibfnamefont {M.}~\bibnamefont {Tsoi}}, \bibinfo {author}
  {\bibfnamefont {T.}~\bibnamefont {Moriyama}}, \bibinfo {author}
  {\bibfnamefont {T.}~\bibnamefont {Ono}},\ and\ \bibinfo {author}
  {\bibfnamefont {Y.}~\bibnamefont {Tserkovnyak}},\ }\bibfield  {title}
  {\bibinfo {title} {Antiferromagnetic spintronics},\ }\href
  {https://doi.org/10.1103/RevModPhys.90.015005} {\bibfield  {journal}
  {\bibinfo  {journal} {Rev. Mod. Phys.}\ }\textbf {\bibinfo {volume} {90}},\
  \bibinfo {pages} {015005} (\bibinfo {year} {2018})}\BibitemShut {NoStop}%
\bibitem [{\citenamefont {Jungwirth}\ \emph {et~al.}(2016)\citenamefont
  {Jungwirth}, \citenamefont {Marti}, \citenamefont {Wadley},\ and\
  \citenamefont {Wunderlich}}]{Jungwirth2016}%
  \BibitemOpen
  \bibfield  {author} {\bibinfo {author} {\bibfnamefont {T.}~\bibnamefont
  {Jungwirth}}, \bibinfo {author} {\bibfnamefont {X.}~\bibnamefont {Marti}},
  \bibinfo {author} {\bibfnamefont {P.}~\bibnamefont {Wadley}},\ and\ \bibinfo
  {author} {\bibfnamefont {J.}~\bibnamefont {Wunderlich}},\ }\bibfield  {title}
  {\bibinfo {title} {Antiferromagnetic spintronics},\ }\href
  {https://doi.org/10.1038/nnano.2016.18} {\bibfield  {journal} {\bibinfo
  {journal} {Nature Nanotechnology}\ }\textbf {\bibinfo {volume} {11}},\
  \bibinfo {pages} {231} (\bibinfo {year} {2016})}\BibitemShut {NoStop}%
\bibitem [{\citenamefont {Gonz\'alez-Hern\'andez}\ \emph
  {et~al.}(2021)\citenamefont {Gonz\'alez-Hern\'andez}, \citenamefont
  {\ifmmode~\check{S}\else \v{S}\fi{}mejkal}, \citenamefont {V\'yborn\'y},
  \citenamefont {Yahagi}, \citenamefont {Sinova}, \citenamefont {Jungwirth},\
  and\ \citenamefont {\ifmmode~\check{Z}\else
  \v{Z}\fi{}elezn\'y}}]{PhysRevLett.126.127701}%
  \BibitemOpen
  \bibfield  {author} {\bibinfo {author} {\bibfnamefont {R.}~\bibnamefont
  {Gonz\'alez-Hern\'andez}}, \bibinfo {author} {\bibfnamefont {L.}~\bibnamefont
  {\ifmmode~\check{S}\else \v{S}\fi{}mejkal}}, \bibinfo {author} {\bibfnamefont
  {K.}~\bibnamefont {V\'yborn\'y}}, \bibinfo {author} {\bibfnamefont
  {Y.}~\bibnamefont {Yahagi}}, \bibinfo {author} {\bibfnamefont
  {J.}~\bibnamefont {Sinova}}, \bibinfo {author} {\bibfnamefont {T.~c.~v.}\
  \bibnamefont {Jungwirth}},\ and\ \bibinfo {author} {\bibfnamefont
  {J.}~\bibnamefont {\ifmmode~\check{Z}\else \v{Z}\fi{}elezn\'y}},\ }\bibfield
  {title} {\bibinfo {title} {Efficient electrical spin splitter based on
  nonrelativistic collinear antiferromagnetism},\ }\href
  {https://doi.org/10.1103/PhysRevLett.126.127701} {\bibfield  {journal}
  {\bibinfo  {journal} {Phys. Rev. Lett.}\ }\textbf {\bibinfo {volume} {126}},\
  \bibinfo {pages} {127701} (\bibinfo {year} {2021})}\BibitemShut {NoStop}%
\bibitem [{\citenamefont {Liu}\ \emph {et~al.}(2022)\citenamefont {Liu},
  \citenamefont {Li}, \citenamefont {Han}, \citenamefont {Wan},\ and\
  \citenamefont {Liu}}]{Liu2022AM}%
  \BibitemOpen
  \bibfield  {author} {\bibinfo {author} {\bibfnamefont {P.}~\bibnamefont
  {Liu}}, \bibinfo {author} {\bibfnamefont {J.}~\bibnamefont {Li}}, \bibinfo
  {author} {\bibfnamefont {J.}~\bibnamefont {Han}}, \bibinfo {author}
  {\bibfnamefont {X.}~\bibnamefont {Wan}},\ and\ \bibinfo {author}
  {\bibfnamefont {Q.}~\bibnamefont {Liu}},\ }\bibfield  {title} {\bibinfo
  {title} {Spin-group symmetry in magnetic materials with negligible spin-orbit
  coupling},\ }\href {https://doi.org/10.1103/PhysRevX.12.021016} {\bibfield
  {journal} {\bibinfo  {journal} {Phys. Rev. X}\ }\textbf {\bibinfo {volume}
  {12}},\ \bibinfo {pages} {021016} (\bibinfo {year} {2022})}\BibitemShut
  {NoStop}%
\bibitem [{\citenamefont {Xiao}\ \emph {et~al.}(2024)\citenamefont {Xiao},
  \citenamefont {Zhao}, \citenamefont {Li}, \citenamefont {Shindou},\ and\
  \citenamefont {Song}}]{Xiao2024SSG}%
  \BibitemOpen
  \bibfield  {author} {\bibinfo {author} {\bibfnamefont {Z.}~\bibnamefont
  {Xiao}}, \bibinfo {author} {\bibfnamefont {J.}~\bibnamefont {Zhao}}, \bibinfo
  {author} {\bibfnamefont {Y.}~\bibnamefont {Li}}, \bibinfo {author}
  {\bibfnamefont {R.}~\bibnamefont {Shindou}},\ and\ \bibinfo {author}
  {\bibfnamefont {Z.-D.}\ \bibnamefont {Song}},\ }\bibfield  {title} {\bibinfo
  {title} {Spin space groups: Full classification and applications},\ }\href
  {https://doi.org/10.1103/PhysRevX.14.031037} {\bibfield  {journal} {\bibinfo
  {journal} {Phys. Rev. X}\ }\textbf {\bibinfo {volume} {14}},\ \bibinfo
  {pages} {031037} (\bibinfo {year} {2024})}\BibitemShut {NoStop}%
\bibitem [{\citenamefont {Jiang}\ \emph {et~al.}(2024)\citenamefont {Jiang},
  \citenamefont {Song}, \citenamefont {Zhu}, \citenamefont {Fang},
  \citenamefont {Weng}, \citenamefont {Liu}, \citenamefont {Yang},\ and\
  \citenamefont {Fang}}]{Yi2024SSG}%
  \BibitemOpen
  \bibfield  {author} {\bibinfo {author} {\bibfnamefont {Y.}~\bibnamefont
  {Jiang}}, \bibinfo {author} {\bibfnamefont {Z.}~\bibnamefont {Song}},
  \bibinfo {author} {\bibfnamefont {T.}~\bibnamefont {Zhu}}, \bibinfo {author}
  {\bibfnamefont {Z.}~\bibnamefont {Fang}}, \bibinfo {author} {\bibfnamefont
  {H.}~\bibnamefont {Weng}}, \bibinfo {author} {\bibfnamefont {Z.-X.}\
  \bibnamefont {Liu}}, \bibinfo {author} {\bibfnamefont {J.}~\bibnamefont
  {Yang}},\ and\ \bibinfo {author} {\bibfnamefont {C.}~\bibnamefont {Fang}},\
  }\bibfield  {title} {\bibinfo {title} {Enumeration of spin-space groups:
  Toward a complete description of symmetries of magnetic orders},\ }\href
  {https://doi.org/10.1103/PhysRevX.14.031039} {\bibfield  {journal} {\bibinfo
  {journal} {Phys. Rev. X}\ }\textbf {\bibinfo {volume} {14}},\ \bibinfo
  {pages} {031039} (\bibinfo {year} {2024})}\BibitemShut {NoStop}%
\bibitem [{\citenamefont {Chen}\ \emph {et~al.}(2024)\citenamefont {Chen},
  \citenamefont {Ren}, \citenamefont {Zhu}, \citenamefont {Yu}, \citenamefont
  {Zhang}, \citenamefont {Liu}, \citenamefont {Li}, \citenamefont {Liu},
  \citenamefont {Li},\ and\ \citenamefont {Liu}}]{Chen2023AM}%
  \BibitemOpen
  \bibfield  {author} {\bibinfo {author} {\bibfnamefont {X.}~\bibnamefont
  {Chen}}, \bibinfo {author} {\bibfnamefont {J.}~\bibnamefont {Ren}}, \bibinfo
  {author} {\bibfnamefont {Y.}~\bibnamefont {Zhu}}, \bibinfo {author}
  {\bibfnamefont {Y.}~\bibnamefont {Yu}}, \bibinfo {author} {\bibfnamefont
  {A.}~\bibnamefont {Zhang}}, \bibinfo {author} {\bibfnamefont
  {P.}~\bibnamefont {Liu}}, \bibinfo {author} {\bibfnamefont {J.}~\bibnamefont
  {Li}}, \bibinfo {author} {\bibfnamefont {Y.}~\bibnamefont {Liu}}, \bibinfo
  {author} {\bibfnamefont {C.}~\bibnamefont {Li}},\ and\ \bibinfo {author}
  {\bibfnamefont {Q.}~\bibnamefont {Liu}},\ }\bibfield  {title} {\bibinfo
  {title} {{Enumeration and Representation Theory of Spin Space Groups}},\
  }\href {https://doi.org/10.1103/PhysRevX.14.031038} {\bibfield  {journal}
  {\bibinfo  {journal} {Phys. Rev. X}\ }\textbf {\bibinfo {volume} {14}},\
  \bibinfo {pages} {031038} (\bibinfo {year} {2024})}\BibitemShut {NoStop}%
\bibitem [{\citenamefont {Pan}\ \emph {et~al.}(2025)\citenamefont {Pan},
  \citenamefont {Zhou}, \citenamefont {Hu}, \citenamefont {Liu}, \citenamefont
  {Zhou}, \citenamefont {Xiao}, \citenamefont {Yang},\ and\ \citenamefont
  {Sun}}]{nn5t-kmln}%
  \BibitemOpen
  \bibfield  {author} {\bibinfo {author} {\bibfnamefont {B.}~\bibnamefont
  {Pan}}, \bibinfo {author} {\bibfnamefont {P.}~\bibnamefont {Zhou}}, \bibinfo
  {author} {\bibfnamefont {Y.}~\bibnamefont {Hu}}, \bibinfo {author}
  {\bibfnamefont {S.}~\bibnamefont {Liu}}, \bibinfo {author} {\bibfnamefont
  {B.}~\bibnamefont {Zhou}}, \bibinfo {author} {\bibfnamefont {H.}~\bibnamefont
  {Xiao}}, \bibinfo {author} {\bibfnamefont {X.}~\bibnamefont {Yang}},\ and\
  \bibinfo {author} {\bibfnamefont {L.}~\bibnamefont {Sun}},\ }\bibfield
  {title} {\bibinfo {title} {Floquet-induced altermagnetic transition in
  ${A}$-type antiferromagnetic bilayers},\ }\href
  {https://doi.org/10.1103/nn5t-kmln} {\bibfield  {journal} {\bibinfo
  {journal} {Phys. Rev. B}\ }\textbf {\bibinfo {volume} {112}},\ \bibinfo
  {pages} {224430} (\bibinfo {year} {2025})}\BibitemShut {NoStop}%
\bibitem [{\citenamefont {Zhu}\ \emph {et~al.}(2026{\natexlab{a}})\citenamefont
  {Zhu}, \citenamefont {Zhou}, \citenamefont {Wang}, \citenamefont {Wei},\ and\
  \citenamefont {Ruan}}]{7ywb-ml2q}%
  \BibitemOpen
  \bibfield  {author} {\bibinfo {author} {\bibfnamefont {T.}~\bibnamefont
  {Zhu}}, \bibinfo {author} {\bibfnamefont {D.}~\bibnamefont {Zhou}}, \bibinfo
  {author} {\bibfnamefont {H.}~\bibnamefont {Wang}}, \bibinfo {author}
  {\bibfnamefont {S.-H.}\ \bibnamefont {Wei}},\ and\ \bibinfo {author}
  {\bibfnamefont {J.}~\bibnamefont {Ruan}},\ }\bibfield  {title} {\bibinfo
  {title} {Floquet odd-parity collinear magnets},\ }\href
  {https://doi.org/10.1103/7ywb-ml2q} {\bibfield  {journal} {\bibinfo
  {journal} {Phys. Rev. Lett.}\ }\textbf {\bibinfo {volume} {136}},\ \bibinfo
  {pages} {126704} (\bibinfo {year} {2026}{\natexlab{a}})}\BibitemShut
  {NoStop}%
\bibitem [{\citenamefont {Huang}\ \emph {et~al.}(2026)\citenamefont {Huang},
  \citenamefont {Qin}, \citenamefont {Zhan}, \citenamefont {Xu}, \citenamefont
  {Ma},\ and\ \citenamefont {Wang}}]{9346-9jpf}%
  \BibitemOpen
  \bibfield  {author} {\bibinfo {author} {\bibfnamefont {S.}~\bibnamefont
  {Huang}}, \bibinfo {author} {\bibfnamefont {Z.}~\bibnamefont {Qin}}, \bibinfo
  {author} {\bibfnamefont {F.}~\bibnamefont {Zhan}}, \bibinfo {author}
  {\bibfnamefont {D.-H.}\ \bibnamefont {Xu}}, \bibinfo {author} {\bibfnamefont
  {D.-S.}\ \bibnamefont {Ma}},\ and\ \bibinfo {author} {\bibfnamefont
  {R.}~\bibnamefont {Wang}},\ }\bibfield  {title} {\bibinfo {title}
  {Light-induced odd-parity magnetism in conventional antiferromagnetism},\
  }\href {https://doi.org/10.1103/9346-9jpf} {\bibfield  {journal} {\bibinfo
  {journal} {Phys. Rev. Lett.}\ }\textbf {\bibinfo {volume} {136}},\ \bibinfo
  {pages} {126703} (\bibinfo {year} {2026})}\BibitemShut {NoStop}%
\bibitem [{\citenamefont {Yarmohammadi}\ \emph {et~al.}(2025)\citenamefont
  {Yarmohammadi}, \citenamefont {Z\"ulicke}, \citenamefont {Berakdar},
  \citenamefont {Linder},\ and\ \citenamefont {Freericks}}]{k3xb-8pts}%
  \BibitemOpen
  \bibfield  {author} {\bibinfo {author} {\bibfnamefont {M.}~\bibnamefont
  {Yarmohammadi}}, \bibinfo {author} {\bibfnamefont {U.}~\bibnamefont
  {Z\"ulicke}}, \bibinfo {author} {\bibfnamefont {J.}~\bibnamefont {Berakdar}},
  \bibinfo {author} {\bibfnamefont {J.}~\bibnamefont {Linder}},\ and\ \bibinfo
  {author} {\bibfnamefont {J.~K.}\ \bibnamefont {Freericks}},\ }\bibfield
  {title} {\bibinfo {title} {Anisotropic light-tailored {RKKY} interaction in
  two-dimensional $d$-wave altermagnets},\ }\href
  {https://doi.org/10.1103/k3xb-8pts} {\bibfield  {journal} {\bibinfo
  {journal} {Phys. Rev. B}\ }\textbf {\bibinfo {volume} {111}},\ \bibinfo
  {pages} {224412} (\bibinfo {year} {2025})}\BibitemShut {NoStop}%
\bibitem [{\citenamefont {Yarmohammadi}\ \emph
  {et~al.}(2026{\natexlab{a}})\citenamefont {Yarmohammadi}, \citenamefont
  {Berritta}, \citenamefont {Bukov}, \citenamefont {\ifmmode~\check{S}\else
  \v{S}\fi{}mejkal}, \citenamefont {Linder},\ and\ \citenamefont
  {Oppeneer}}]{xt23-9pnv}%
  \BibitemOpen
  \bibfield  {author} {\bibinfo {author} {\bibfnamefont {M.}~\bibnamefont
  {Yarmohammadi}}, \bibinfo {author} {\bibfnamefont {M.}~\bibnamefont
  {Berritta}}, \bibinfo {author} {\bibfnamefont {M.}~\bibnamefont {Bukov}},
  \bibinfo {author} {\bibfnamefont {L.}~\bibnamefont {\ifmmode~\check{S}\else
  \v{S}\fi{}mejkal}}, \bibinfo {author} {\bibfnamefont {J.}~\bibnamefont
  {Linder}},\ and\ \bibinfo {author} {\bibfnamefont {P.~M.}\ \bibnamefont
  {Oppeneer}},\ }\bibfield  {title} {\bibinfo {title} {Spin polarization
  engineering in $d$-wave altermagnets},\ }\href
  {https://doi.org/10.1103/xt23-9pnv} {\bibfield  {journal} {\bibinfo
  {journal} {Phys. Rev. B}\ }\textbf {\bibinfo {volume} {113}},\ \bibinfo
  {pages} {L060403} (\bibinfo {year} {2026}{\natexlab{a}})}\BibitemShut
  {NoStop}%
\bibitem [{\citenamefont {Zou}\ \emph {et~al.}(2026)\citenamefont {Zou},
  \citenamefont {Shin}, \citenamefont {Huang}, \citenamefont {Zang},
  \citenamefont {Dai}, \citenamefont {Niu}, \citenamefont {Kang},\ and\
  \citenamefont
  {Myung}}]{zou2026floquetengineeredoddparityaltermagnetichigherorder}%
  \BibitemOpen
  \bibfield  {author} {\bibinfo {author} {\bibfnamefont {X.}~\bibnamefont
  {Zou}}, \bibinfo {author} {\bibfnamefont {H.~S.}\ \bibnamefont {Shin}},
  \bibinfo {author} {\bibfnamefont {B.}~\bibnamefont {Huang}}, \bibinfo
  {author} {\bibfnamefont {Y.}~\bibnamefont {Zang}}, \bibinfo {author}
  {\bibfnamefont {Y.}~\bibnamefont {Dai}}, \bibinfo {author} {\bibfnamefont
  {C.}~\bibnamefont {Niu}}, \bibinfo {author} {\bibfnamefont {C.-J.}\
  \bibnamefont {Kang}},\ and\ \bibinfo {author} {\bibfnamefont {C.~W.}\
  \bibnamefont {Myung}},\ }\bibfield  {title} {\bibinfo {title}
  {Floquet-engineered odd-parity altermagnetic higher-order topology in a
  two-dimensional antiferromagnet {C}r$_2${CH}$_2$},\ }\href
  {https://arxiv.org/abs/2605.19184} {\  (\bibinfo {year} {2026})},\ \Eprint
  {https://arxiv.org/abs/2605.19184} {arXiv:2605.19184} \BibitemShut {NoStop}%
\bibitem [{\citenamefont {Yarmohammadi}\ \emph
  {et~al.}(2026{\natexlab{b}})\citenamefont {Yarmohammadi}, \citenamefont
  {Gunnink}, \citenamefont {Sinova},\ and\ \citenamefont
  {Freericks}}]{yarmohammadi2026floquetinducedanisotropicmagnetoresistanceanomalous}%
  \BibitemOpen
  \bibfield  {author} {\bibinfo {author} {\bibfnamefont {M.}~\bibnamefont
  {Yarmohammadi}}, \bibinfo {author} {\bibfnamefont {P.~M.}\ \bibnamefont
  {Gunnink}}, \bibinfo {author} {\bibfnamefont {J.}~\bibnamefont {Sinova}},\
  and\ \bibinfo {author} {\bibfnamefont {J.~K.}\ \bibnamefont {Freericks}},\
  }\bibfield  {title} {\bibinfo {title} {Floquet-induced anisotropic
  magnetoresistance and anomalous {H}all effect in 2{D} $d$-wave altermagnets
  with {R}ashba spin-orbit coupling},\ }\href
  {https://arxiv.org/abs/2606.21281} {\  (\bibinfo {year}
  {2026}{\natexlab{b}})},\ \Eprint {https://arxiv.org/abs/2606.21281}
  {arXiv:2606.21281} \BibitemShut {NoStop}%
\bibitem [{\citenamefont {Yarmohammadi}\ \emph
  {et~al.}(2026{\natexlab{c}})\citenamefont {Yarmohammadi}, \citenamefont {Jo},
  \citenamefont {Berritta}, \citenamefont {Šmejkal}, \citenamefont
  {Freericks},\ and\ \citenamefont
  {Oppeneer}}]{yarmohammadi2026giantperpendicularedelsteinpolarization}%
  \BibitemOpen
  \bibfield  {author} {\bibinfo {author} {\bibfnamefont {M.}~\bibnamefont
  {Yarmohammadi}}, \bibinfo {author} {\bibfnamefont {D.}~\bibnamefont {Jo}},
  \bibinfo {author} {\bibfnamefont {M.}~\bibnamefont {Berritta}}, \bibinfo
  {author} {\bibfnamefont {L.}~\bibnamefont {Šmejkal}}, \bibinfo {author}
  {\bibfnamefont {J.~K.}\ \bibnamefont {Freericks}},\ and\ \bibinfo {author}
  {\bibfnamefont {P.~M.}\ \bibnamefont {Oppeneer}},\ }\bibfield  {title}
  {\bibinfo {title} {Giant perpendicular {E}delstein polarization in 2{D}
  compensated magnets via bichromatic {F}loquet driving},\ }\href
  {https://arxiv.org/abs/2606.31867} {\  (\bibinfo {year}
  {2026}{\natexlab{c}})},\ \Eprint {https://arxiv.org/abs/2606.31867}
  {arXiv:2606.31867} \BibitemShut {NoStop}%
\bibitem [{\citenamefont {Yarmohammadi}\ \emph
  {et~al.}(2026{\natexlab{d}})\citenamefont {Yarmohammadi}, \citenamefont
  {Fu},\ and\ \citenamefont
  {Freericks}}]{yarmohammadi2026efficienttwocolorfloquetcontrol}%
  \BibitemOpen
  \bibfield  {author} {\bibinfo {author} {\bibfnamefont {M.}~\bibnamefont
  {Yarmohammadi}}, \bibinfo {author} {\bibfnamefont {P.-H.}\ \bibnamefont
  {Fu}},\ and\ \bibinfo {author} {\bibfnamefont {J.~K.}\ \bibnamefont
  {Freericks}},\ }\bibfield  {title} {\bibinfo {title} {Efficient two-color
  floquet control of the {RKKY} interaction in altermagnets},\ }\href
  {https://arxiv.org/abs/2602.20862} {\  (\bibinfo {year}
  {2026}{\natexlab{d}})},\ \Eprint {https://arxiv.org/abs/2602.20862}
  {arXiv:2602.20862} \BibitemShut {NoStop}%
\bibitem [{\citenamefont {Yu}(2026)}]{yu2026tunableoddparityspinsplittings}%
  \BibitemOpen
  \bibfield  {author} {\bibinfo {author} {\bibfnamefont {Y.}~\bibnamefont
  {Yu}},\ }\bibfield  {title} {\bibinfo {title} {Tunable odd-parity spin
  splittings in altermagnets},\ }\href {https://arxiv.org/abs/2605.03026} {\
  (\bibinfo {year} {2026})},\ \Eprint {https://arxiv.org/abs/2605.03026}
  {arXiv:2605.03026} \BibitemShut {NoStop}%
\bibitem [{\citenamefont {Li}\ \emph {et~al.}(2026)\citenamefont {Li},
  \citenamefont {Shao},\ and\ \citenamefont {Kovalev}}]{Li2025AM7}%
  \BibitemOpen
  \bibfield  {author} {\bibinfo {author} {\bibfnamefont {B.}~\bibnamefont
  {Li}}, \bibinfo {author} {\bibfnamefont {D.-F.}\ \bibnamefont {Shao}},\ and\
  \bibinfo {author} {\bibfnamefont {A.~A.}\ \bibnamefont {Kovalev}},\
  }\bibfield  {title} {\bibinfo {title} {Floquet spin splitting and spin
  generation in antiferromagnets},\ }\href {https://doi.org/10.1103/xzm1-l6yf}
  {\bibfield  {journal} {\bibinfo  {journal} {Phys. Rev. Lett.}\ }\textbf
  {\bibinfo {volume} {136}},\ \bibinfo {pages} {166701} (\bibinfo {year}
  {2026})}\BibitemShut {NoStop}%
\bibitem [{\citenamefont {Liu}\ \emph {et~al.}(2026{\natexlab{a}})\citenamefont
  {Liu}, \citenamefont {Zhuang}, \citenamefont {Zhu}, \citenamefont {Wu},\ and\
  \citenamefont {Yan}}]{liu2025AM10}%
  \BibitemOpen
  \bibfield  {author} {\bibinfo {author} {\bibfnamefont {D.}~\bibnamefont
  {Liu}}, \bibinfo {author} {\bibfnamefont {Z.-Y.}\ \bibnamefont {Zhuang}},
  \bibinfo {author} {\bibfnamefont {D.}~\bibnamefont {Zhu}}, \bibinfo {author}
  {\bibfnamefont {Z.}~\bibnamefont {Wu}},\ and\ \bibinfo {author}
  {\bibfnamefont {Z.}~\bibnamefont {Yan}},\ }\bibfield  {title} {\bibinfo
  {title} {Light-induced odd-parity altermagnets on dimerized lattices},\
  }\href {https://doi.org/10.1103/wnqs-3djt} {\bibfield  {journal} {\bibinfo
  {journal} {Phys. Rev. B}\ }\textbf {\bibinfo {volume} {113}},\ \bibinfo
  {pages} {L060409} (\bibinfo {year} {2026}{\natexlab{a}})}\BibitemShut
  {NoStop}%
\bibitem [{\citenamefont {Liu}\ \emph {et~al.}(2026{\natexlab{b}})\citenamefont
  {Liu}, \citenamefont {Zhuang}, \citenamefont {Zhu}, \citenamefont {Wu},\ and\
  \citenamefont {Yan}}]{liu2026AM}%
  \BibitemOpen
  \bibfield  {author} {\bibinfo {author} {\bibfnamefont {D.}~\bibnamefont
  {Liu}}, \bibinfo {author} {\bibfnamefont {Z.-Y.}\ \bibnamefont {Zhuang}},
  \bibinfo {author} {\bibfnamefont {D.}~\bibnamefont {Zhu}}, \bibinfo {author}
  {\bibfnamefont {Z.}~\bibnamefont {Wu}},\ and\ \bibinfo {author}
  {\bibfnamefont {Z.}~\bibnamefont {Yan}},\ }\bibfield  {title} {\bibinfo
  {title} {Nonrelativistic spin-orbit-coupling effects in odd-parity coplanar
  magnets},\ }\href {https://arxiv.org/abs/2606.23806} {\  (\bibinfo {year}
  {2026}{\natexlab{b}})},\ \Eprint {https://arxiv.org/abs/2606.23806}
  {arXiv:2606.23806} \BibitemShut {NoStop}%
\bibitem [{\citenamefont {Zhu}\ \emph {et~al.}(2026{\natexlab{b}})\citenamefont
  {Zhu}, \citenamefont {Liu}, \citenamefont {Zhuang}, \citenamefont {Wu},\ and\
  \citenamefont {Yan}}]{zhu2026BCAFM}%
  \BibitemOpen
  \bibfield  {author} {\bibinfo {author} {\bibfnamefont {D.}~\bibnamefont
  {Zhu}}, \bibinfo {author} {\bibfnamefont {D.}~\bibnamefont {Liu}}, \bibinfo
  {author} {\bibfnamefont {Z.-Y.}\ \bibnamefont {Zhuang}}, \bibinfo {author}
  {\bibfnamefont {Z.}~\bibnamefont {Wu}},\ and\ \bibinfo {author}
  {\bibfnamefont {Z.}~\bibnamefont {Yan}},\ }\bibfield  {title} {\bibinfo
  {title} {Light-induced even-parity unidirectional spin splitting in coplanar
  antiferromagnets},\ }\href {https://arxiv.org/abs/2601.03358} {\  (\bibinfo
  {year} {2026}{\natexlab{b}})},\ \Eprint {https://arxiv.org/abs/2601.03358}
  {arXiv:2601.03358} \BibitemShut {NoStop}%
\bibitem [{\citenamefont
  {Yarmohammadi}(2026)}]{yarmohammadi2026chirpedfloquetlineardrives}%
  \BibitemOpen
  \bibfield  {author} {\bibinfo {author} {\bibfnamefont {M.}~\bibnamefont
  {Yarmohammadi}},\ }\bibfield  {title} {\bibinfo {title} {Chirped floquet
  linear drives activate forbidden charge-to-spin conversions in {R}ashba
  two-dimensional electron gases},\ }\href {https://arxiv.org/abs/2607.04946}
  {\  (\bibinfo {year} {2026})},\ \Eprint {https://arxiv.org/abs/2607.04946}
  {arXiv:2607.04946} \BibitemShut {NoStop}%
\bibitem [{SM()}]{SM}%
  \BibitemOpen
  \href@noop {} {\bibinfo  {journal} {See Supplemental Material at [URL will be
  inserted by publisher] for the detailed derivation of the effective
  Hamiltonian for the bichromatically driven coplanar antiferromagnet,
  including high-temperature expansion and a comparative analysis of the
  fundamental and higher-order harmonic driving regimes}\ }\BibitemShut
  {NoStop}%
\bibitem [{\citenamefont {Hejazi}\ \emph {et~al.}(2020)\citenamefont {Hejazi},
  \citenamefont {Luo},\ and\ \citenamefont {Balents}}]{hejazi2020noncollinear}%
  \BibitemOpen
\bibfield  {journal} {  }\bibfield  {author} {\bibinfo {author} {\bibfnamefont
  {K.}~\bibnamefont {Hejazi}}, \bibinfo {author} {\bibfnamefont {Z.-X.}\
  \bibnamefont {Luo}},\ and\ \bibinfo {author} {\bibfnamefont {L.}~\bibnamefont
  {Balents}},\ }\bibfield  {title} {\bibinfo {title} {Noncollinear phases in
  moir{\'e} magnets},\ }\href {https://doi.org/10.1073/pnas.2000347117}
  {\bibfield  {journal} {\bibinfo  {journal} {Proceedings of the National
  Academy of Sciences}\ }\textbf {\bibinfo {volume} {117}},\ \bibinfo {pages}
  {10721} (\bibinfo {year} {2020})}\BibitemShut {NoStop}%
\bibitem [{\citenamefont {Healey}\ \emph {et~al.}(2025)\citenamefont {Healey},
  \citenamefont {Tan}, \citenamefont {Gross}, \citenamefont {Scholten},
  \citenamefont {Xing}, \citenamefont {Chica}, \citenamefont {Johnson},
  \citenamefont {Poggio}, \citenamefont {Ziebel}, \citenamefont {Roy},
  \citenamefont {Tetienne},\ and\ \citenamefont
  {Broadway}}]{robertson2025imaging}%
  \BibitemOpen
  \bibfield  {author} {\bibinfo {author} {\bibfnamefont {A.~J.}\ \bibnamefont
  {Healey}}, \bibinfo {author} {\bibfnamefont {C.}~\bibnamefont {Tan}},
  \bibinfo {author} {\bibfnamefont {B.}~\bibnamefont {Gross}}, \bibinfo
  {author} {\bibfnamefont {S.~C.}\ \bibnamefont {Scholten}}, \bibinfo {author}
  {\bibfnamefont {K.}~\bibnamefont {Xing}}, \bibinfo {author} {\bibfnamefont
  {D.~G.}\ \bibnamefont {Chica}}, \bibinfo {author} {\bibfnamefont {B.~C.}\
  \bibnamefont {Johnson}}, \bibinfo {author} {\bibfnamefont {M.}~\bibnamefont
  {Poggio}}, \bibinfo {author} {\bibfnamefont {M.~E.}\ \bibnamefont {Ziebel}},
  \bibinfo {author} {\bibfnamefont {X.}~\bibnamefont {Roy}}, \bibinfo {author}
  {\bibfnamefont {J.-P.}\ \bibnamefont {Tetienne}},\ and\ \bibinfo {author}
  {\bibfnamefont {D.~A.}\ \bibnamefont {Broadway}},\ }\bibfield  {title}
  {\bibinfo {title} {Imaging magnetic switching in orthogonally twisted stacks
  of a van der {W}aals antiferromagnet},\ }\href
  {https://doi.org/10.1021/acsnano.5c12297} {\bibfield  {journal} {\bibinfo
  {journal} {ACS Nano}\ }\textbf {\bibinfo {volume} {19}},\ \bibinfo {pages}
  {42140} (\bibinfo {year} {2025})}\BibitemShut {NoStop}%
\bibitem [{\citenamefont {Liu}\ \emph {et~al.}(2024)\citenamefont {Liu},
  \citenamefont {Yu},\ and\ \citenamefont {Liu}}]{LiuTwistAM2024}%
  \BibitemOpen
  \bibfield  {author} {\bibinfo {author} {\bibfnamefont {Y.}~\bibnamefont
  {Liu}}, \bibinfo {author} {\bibfnamefont {J.}~\bibnamefont {Yu}},\ and\
  \bibinfo {author} {\bibfnamefont {C.-C.}\ \bibnamefont {Liu}},\ }\bibfield
  {title} {\bibinfo {title} {Twisted magnetic van der {W}aals bilayers: An
  ideal platform for altermagnetism},\ }\href
  {https://doi.org/10.1103/PhysRevLett.133.206702} {\bibfield  {journal}
  {\bibinfo  {journal} {Phys. Rev. Lett.}\ }\textbf {\bibinfo {volume} {133}},\
  \bibinfo {pages} {206702} (\bibinfo {year} {2024})}\BibitemShut {NoStop}%
\bibitem [{\citenamefont {Bai}\ \emph {et~al.}(2025)\citenamefont {Bai},
  \citenamefont {Zhang}, \citenamefont {Feng},\ and\ \citenamefont
  {Yao}}]{BaiTypeIVAFM2025}%
  \BibitemOpen
  \bibfield  {author} {\bibinfo {author} {\bibfnamefont {L.}~\bibnamefont
  {Bai}}, \bibinfo {author} {\bibfnamefont {R.-W.}\ \bibnamefont {Zhang}},
  \bibinfo {author} {\bibfnamefont {W.}~\bibnamefont {Feng}},\ and\ \bibinfo
  {author} {\bibfnamefont {Y.}~\bibnamefont {Yao}},\ }\bibfield  {title}
  {\bibinfo {title} {Anomalous {H}all effect in type {IV} 2{D} collinear
  magnets},\ }\href {https://doi.org/10.1103/rn1l-d6cq} {\bibfield  {journal}
  {\bibinfo  {journal} {Phys. Rev. Lett.}\ }\textbf {\bibinfo {volume} {135}},\
  \bibinfo {pages} {036702} (\bibinfo {year} {2025})}\BibitemShut {NoStop}%
\bibitem [{\citenamefont {Gibert}\ \emph {et~al.}(2015)\citenamefont {Gibert},
  \citenamefont {Viret}, \citenamefont {Torres-Pardo}, \citenamefont
  {Piamonteze}, \citenamefont {Zubko}, \citenamefont {Jaouen}, \citenamefont
  {Tonnerre}, \citenamefont {Mougin}, \citenamefont {Fowlie}, \citenamefont
  {Catalano}, \citenamefont {Gloter}, \citenamefont {St{\'e}phan},\ and\
  \citenamefont {Triscone}}]{gibert2015interfacial}%
  \BibitemOpen
  \bibfield  {author} {\bibinfo {author} {\bibfnamefont {M.}~\bibnamefont
  {Gibert}}, \bibinfo {author} {\bibfnamefont {M.}~\bibnamefont {Viret}},
  \bibinfo {author} {\bibfnamefont {A.}~\bibnamefont {Torres-Pardo}}, \bibinfo
  {author} {\bibfnamefont {C.}~\bibnamefont {Piamonteze}}, \bibinfo {author}
  {\bibfnamefont {P.}~\bibnamefont {Zubko}}, \bibinfo {author} {\bibfnamefont
  {N.}~\bibnamefont {Jaouen}}, \bibinfo {author} {\bibfnamefont {J.-M.}\
  \bibnamefont {Tonnerre}}, \bibinfo {author} {\bibfnamefont {A.}~\bibnamefont
  {Mougin}}, \bibinfo {author} {\bibfnamefont {J.}~\bibnamefont {Fowlie}},
  \bibinfo {author} {\bibfnamefont {S.}~\bibnamefont {Catalano}}, \bibinfo
  {author} {\bibfnamefont {A.}~\bibnamefont {Gloter}}, \bibinfo {author}
  {\bibfnamefont {O.}~\bibnamefont {St{\'e}phan}},\ and\ \bibinfo {author}
  {\bibfnamefont {J.-M.}\ \bibnamefont {Triscone}},\ }\bibfield  {title}
  {\bibinfo {title} {Interfacial control of magnetic properties at
  {L}a{M}n{O}$_3$/{L}a{N}i{O}$_3$ interfaces},\ }\href
  {https://doi.org/10.1021/acs.nanolett.5b02720} {\bibfield  {journal}
  {\bibinfo  {journal} {Nano Letters}\ }\textbf {\bibinfo {volume} {15}},\
  \bibinfo {pages} {7355} (\bibinfo {year} {2015})}\BibitemShut {NoStop}%
\bibitem [{\citenamefont {Mudd}\ \emph {et~al.}(2013)\citenamefont {Mudd},
  \citenamefont {Svatek}, \citenamefont {Ren}, \citenamefont {Patanè},
  \citenamefont {Makarovsky}, \citenamefont {Eaves}, \citenamefont {Beton},
  \citenamefont {Kovalyuk}, \citenamefont {Lashkarev}, \citenamefont
  {Kudrynskyi},\ and\ \citenamefont
  {Dmitriev}}]{https://doi.org/10.1002/adma.201302616}%
  \BibitemOpen
  \bibfield  {author} {\bibinfo {author} {\bibfnamefont {G.~W.}\ \bibnamefont
  {Mudd}}, \bibinfo {author} {\bibfnamefont {S.~A.}\ \bibnamefont {Svatek}},
  \bibinfo {author} {\bibfnamefont {T.}~\bibnamefont {Ren}}, \bibinfo {author}
  {\bibfnamefont {A.}~\bibnamefont {Patanè}}, \bibinfo {author} {\bibfnamefont
  {O.}~\bibnamefont {Makarovsky}}, \bibinfo {author} {\bibfnamefont
  {L.}~\bibnamefont {Eaves}}, \bibinfo {author} {\bibfnamefont {P.~H.}\
  \bibnamefont {Beton}}, \bibinfo {author} {\bibfnamefont {Z.~D.}\ \bibnamefont
  {Kovalyuk}}, \bibinfo {author} {\bibfnamefont {G.~V.}\ \bibnamefont
  {Lashkarev}}, \bibinfo {author} {\bibfnamefont {Z.~R.}\ \bibnamefont
  {Kudrynskyi}},\ and\ \bibinfo {author} {\bibfnamefont {A.~I.}\ \bibnamefont
  {Dmitriev}},\ }\bibfield  {title} {\bibinfo {title} {Tuning the bandgap of
  exfoliated {I}n{S}e nanosheets by quantum confinement},\ }\href
  {https://doi.org/https://doi.org/10.1002/adma.201302616} {\bibfield
  {journal} {\bibinfo  {journal} {Advanced Materials}\ }\textbf {\bibinfo
  {volume} {25}},\ \bibinfo {pages} {5714} (\bibinfo {year}
  {2013})}\BibitemShut {NoStop}%
\bibitem [{\citenamefont {Sobota}\ \emph {et~al.}(2012)\citenamefont {Sobota},
  \citenamefont {Yang}, \citenamefont {Analytis}, \citenamefont {Chen},
  \citenamefont {Fisher}, \citenamefont {Kirchmann},\ and\ \citenamefont
  {Shen}}]{PhysRevLett.108.117403}%
  \BibitemOpen
  \bibfield  {author} {\bibinfo {author} {\bibfnamefont {J.~A.}\ \bibnamefont
  {Sobota}}, \bibinfo {author} {\bibfnamefont {S.}~\bibnamefont {Yang}},
  \bibinfo {author} {\bibfnamefont {J.~G.}\ \bibnamefont {Analytis}}, \bibinfo
  {author} {\bibfnamefont {Y.~L.}\ \bibnamefont {Chen}}, \bibinfo {author}
  {\bibfnamefont {I.~R.}\ \bibnamefont {Fisher}}, \bibinfo {author}
  {\bibfnamefont {P.~S.}\ \bibnamefont {Kirchmann}},\ and\ \bibinfo {author}
  {\bibfnamefont {Z.-X.}\ \bibnamefont {Shen}},\ }\bibfield  {title} {\bibinfo
  {title} {Ultrafast optical excitation of a persistent surface-state
  population in the topological insulator {B}i$_{2}${S}e$_{3}$},\ }\href
  {https://doi.org/10.1103/PhysRevLett.108.117403} {\bibfield  {journal}
  {\bibinfo  {journal} {Phys. Rev. Lett.}\ }\textbf {\bibinfo {volume} {108}},\
  \bibinfo {pages} {117403} (\bibinfo {year} {2012})}\BibitemShut {NoStop}%
\bibitem [{\citenamefont {Sobota}\ \emph {et~al.}(2021)\citenamefont {Sobota},
  \citenamefont {He},\ and\ \citenamefont {Shen}}]{RevModPhys.93.025006}%
  \BibitemOpen
  \bibfield  {author} {\bibinfo {author} {\bibfnamefont {J.~A.}\ \bibnamefont
  {Sobota}}, \bibinfo {author} {\bibfnamefont {Y.}~\bibnamefont {He}},\ and\
  \bibinfo {author} {\bibfnamefont {Z.-X.}\ \bibnamefont {Shen}},\ }\bibfield
  {title} {\bibinfo {title} {Angle-resolved photoemission studies of quantum
  materials},\ }\href {https://doi.org/10.1103/RevModPhys.93.025006} {\bibfield
   {journal} {\bibinfo  {journal} {Rev. Mod. Phys.}\ }\textbf {\bibinfo
  {volume} {93}},\ \bibinfo {pages} {025006} (\bibinfo {year}
  {2021})}\BibitemShut {NoStop}%
\bibitem [{\citenamefont {Basov}\ \emph {et~al.}(2017)\citenamefont {Basov},
  \citenamefont {Averitt},\ and\ \citenamefont {Hsieh}}]{Basov2017}%
  \BibitemOpen
  \bibfield  {author} {\bibinfo {author} {\bibfnamefont {D.~N.}\ \bibnamefont
  {Basov}}, \bibinfo {author} {\bibfnamefont {R.~D.}\ \bibnamefont {Averitt}},\
  and\ \bibinfo {author} {\bibfnamefont {D.}~\bibnamefont {Hsieh}},\ }\bibfield
   {title} {\bibinfo {title} {Towards properties on demand in quantum
  materials},\ }\href {https://doi.org/10.1038/nmat5017} {\bibfield  {journal}
  {\bibinfo  {journal} {Nature Materials}\ }\textbf {\bibinfo {volume} {16}},\
  \bibinfo {pages} {1077} (\bibinfo {year} {2017})}\BibitemShut {NoStop}%
\bibitem [{\citenamefont {Hajlaoui}\ \emph {et~al.}(2014)\citenamefont
  {Hajlaoui}, \citenamefont {Papalazarou}, \citenamefont {Mauchain},
  \citenamefont {Perfetti}, \citenamefont {Taleb-Ibrahimi}, \citenamefont
  {Navarin}, \citenamefont {Monteverde}, \citenamefont {Auban-Senzier},
  \citenamefont {Pasquier}, \citenamefont {Moisan}, \citenamefont {Boschetto},
  \citenamefont {Neupane}, \citenamefont {Hasan}, \citenamefont {Durakiewicz},
  \citenamefont {Jiang}, \citenamefont {Xu}, \citenamefont {Miotkowski},
  \citenamefont {Chen}, \citenamefont {Jia}, \citenamefont {Ji}, \citenamefont
  {Cava},\ and\ \citenamefont {Marsi}}]{Hajlaoui2014}%
  \BibitemOpen
  \bibfield  {author} {\bibinfo {author} {\bibfnamefont {M.}~\bibnamefont
  {Hajlaoui}}, \bibinfo {author} {\bibfnamefont {E.}~\bibnamefont
  {Papalazarou}}, \bibinfo {author} {\bibfnamefont {J.}~\bibnamefont
  {Mauchain}}, \bibinfo {author} {\bibfnamefont {L.}~\bibnamefont {Perfetti}},
  \bibinfo {author} {\bibfnamefont {A.}~\bibnamefont {Taleb-Ibrahimi}},
  \bibinfo {author} {\bibfnamefont {F.}~\bibnamefont {Navarin}}, \bibinfo
  {author} {\bibfnamefont {M.}~\bibnamefont {Monteverde}}, \bibinfo {author}
  {\bibfnamefont {P.}~\bibnamefont {Auban-Senzier}}, \bibinfo {author}
  {\bibfnamefont {C.~R.}\ \bibnamefont {Pasquier}}, \bibinfo {author}
  {\bibfnamefont {N.}~\bibnamefont {Moisan}}, \bibinfo {author} {\bibfnamefont
  {D.}~\bibnamefont {Boschetto}}, \bibinfo {author} {\bibfnamefont
  {M.}~\bibnamefont {Neupane}}, \bibinfo {author} {\bibfnamefont {M.~Z.}\
  \bibnamefont {Hasan}}, \bibinfo {author} {\bibfnamefont {T.}~\bibnamefont
  {Durakiewicz}}, \bibinfo {author} {\bibfnamefont {Z.}~\bibnamefont {Jiang}},
  \bibinfo {author} {\bibfnamefont {Y.}~\bibnamefont {Xu}}, \bibinfo {author}
  {\bibfnamefont {I.}~\bibnamefont {Miotkowski}}, \bibinfo {author}
  {\bibfnamefont {Y.~P.}\ \bibnamefont {Chen}}, \bibinfo {author}
  {\bibfnamefont {S.}~\bibnamefont {Jia}}, \bibinfo {author} {\bibfnamefont
  {H.~W.}\ \bibnamefont {Ji}}, \bibinfo {author} {\bibfnamefont {R.~J.}\
  \bibnamefont {Cava}},\ and\ \bibinfo {author} {\bibfnamefont
  {M.}~\bibnamefont {Marsi}},\ }\bibfield  {title} {\bibinfo {title} {Tuning a
  {S}chottky barrier in a photoexcited topological insulator with transient
  {D}irac cone electron-hole asymmetry},\ }\href
  {https://doi.org/10.1038/ncomms4003} {\bibfield  {journal} {\bibinfo
  {journal} {Nature Communications}\ }\textbf {\bibinfo {volume} {5}},\
  \bibinfo {pages} {3003} (\bibinfo {year} {2014})}\BibitemShut {NoStop}%
\bibitem [{\citenamefont {Kitagawa}\ \emph {et~al.}(2011)\citenamefont
  {Kitagawa}, \citenamefont {Oka}, \citenamefont {Brataas}, \citenamefont
  {Fu},\ and\ \citenamefont {Demler}}]{Kitagawa2011EffectiveH}%
  \BibitemOpen
  \bibfield  {author} {\bibinfo {author} {\bibfnamefont {T.}~\bibnamefont
  {Kitagawa}}, \bibinfo {author} {\bibfnamefont {T.}~\bibnamefont {Oka}},
  \bibinfo {author} {\bibfnamefont {A.}~\bibnamefont {Brataas}}, \bibinfo
  {author} {\bibfnamefont {L.}~\bibnamefont {Fu}},\ and\ \bibinfo {author}
  {\bibfnamefont {E.}~\bibnamefont {Demler}},\ }\bibfield  {title} {\bibinfo
  {title} {Transport properties of nonequilibrium systems under the application
  of light: Photoinduced quantum {H}all insulators without {L}andau levels},\
  }\href {https://doi.org/10.1103/PhysRevB.84.235108} {\bibfield  {journal}
  {\bibinfo  {journal} {Phys. Rev. B}\ }\textbf {\bibinfo {volume} {84}},\
  \bibinfo {pages} {235108} (\bibinfo {year} {2011})}\BibitemShut {NoStop}%
\bibitem [{\citenamefont {Goldman}\ and\ \citenamefont
  {Dalibard}(2014)}]{Goldman2014EffectiveH}%
  \BibitemOpen
  \bibfield  {author} {\bibinfo {author} {\bibfnamefont {N.}~\bibnamefont
  {Goldman}}\ and\ \bibinfo {author} {\bibfnamefont {J.}~\bibnamefont
  {Dalibard}},\ }\bibfield  {title} {\bibinfo {title} {Periodically driven
  quantum systems: Effective {H}amiltonians and engineered gauge fields},\
  }\href {https://doi.org/10.1103/PhysRevX.4.031027} {\bibfield  {journal}
  {\bibinfo  {journal} {Phys. Rev. X}\ }\textbf {\bibinfo {volume} {4}},\
  \bibinfo {pages} {031027} (\bibinfo {year} {2014})}\BibitemShut {NoStop}%
\bibitem [{\citenamefont {Eckardt}\ and\ \citenamefont
  {Anisimovas}(2015)}]{eckardt2015EffectiveH}%
  \BibitemOpen
  \bibfield  {author} {\bibinfo {author} {\bibfnamefont {A.}~\bibnamefont
  {Eckardt}}\ and\ \bibinfo {author} {\bibfnamefont {E.}~\bibnamefont
  {Anisimovas}},\ }\bibfield  {title} {\bibinfo {title} {High-frequency
  approximation for periodically driven quantum systems from a {F}loquet-space
  perspective},\ }\href@noop {} {\bibfield  {journal} {\bibinfo  {journal} {New
  journal of physics}\ }\textbf {\bibinfo {volume} {17}},\ \bibinfo {pages}
  {093039} (\bibinfo {year} {2015})}\BibitemShut {NoStop}%
\bibitem [{\citenamefont {Hayami}\ \emph
  {et~al.}(2020{\natexlab{a}})\citenamefont {Hayami}, \citenamefont {Yanagi},\
  and\ \citenamefont {Kusunose}}]{Hayami2020AM1}%
  \BibitemOpen
  \bibfield  {author} {\bibinfo {author} {\bibfnamefont {S.}~\bibnamefont
  {Hayami}}, \bibinfo {author} {\bibfnamefont {Y.}~\bibnamefont {Yanagi}},\
  and\ \bibinfo {author} {\bibfnamefont {H.}~\bibnamefont {Kusunose}},\
  }\bibfield  {title} {\bibinfo {title} {Bottom-up design of spin-split and
  reshaped electronic band structures in antiferromagnets without spin-orbit
  coupling: Procedure on the basis of augmented multipoles},\ }\href
  {https://doi.org/10.1103/PhysRevB.102.144441} {\bibfield  {journal} {\bibinfo
   {journal} {Phys. Rev. B}\ }\textbf {\bibinfo {volume} {102}},\ \bibinfo
  {pages} {144441} (\bibinfo {year} {2020}{\natexlab{a}})}\BibitemShut
  {NoStop}%
\bibitem [{\citenamefont {Hayami}\ \emph
  {et~al.}(2020{\natexlab{b}})\citenamefont {Hayami}, \citenamefont {Yanagi},\
  and\ \citenamefont {Kusunose}}]{Hayami2020AM2}%
  \BibitemOpen
  \bibfield  {author} {\bibinfo {author} {\bibfnamefont {S.}~\bibnamefont
  {Hayami}}, \bibinfo {author} {\bibfnamefont {Y.}~\bibnamefont {Yanagi}},\
  and\ \bibinfo {author} {\bibfnamefont {H.}~\bibnamefont {Kusunose}},\
  }\bibfield  {title} {\bibinfo {title} {Spontaneous antisymmetric spin
  splitting in noncollinear antiferromagnets without spin-orbit coupling},\
  }\href {https://doi.org/10.1103/PhysRevB.101.220403} {\bibfield  {journal}
  {\bibinfo  {journal} {Phys. Rev. B}\ }\textbf {\bibinfo {volume} {101}},\
  \bibinfo {pages} {220403} (\bibinfo {year} {2020}{\natexlab{b}})}\BibitemShut
  {NoStop}%
\bibitem [{\citenamefont {Dehghani}\ \emph {et~al.}(2015)\citenamefont
  {Dehghani}, \citenamefont {Oka},\ and\ \citenamefont
  {Mitra}}]{Dehghani2015quench}%
  \BibitemOpen
  \bibfield  {author} {\bibinfo {author} {\bibfnamefont {H.}~\bibnamefont
  {Dehghani}}, \bibinfo {author} {\bibfnamefont {T.}~\bibnamefont {Oka}},\ and\
  \bibinfo {author} {\bibfnamefont {A.}~\bibnamefont {Mitra}},\ }\bibfield
  {title} {\bibinfo {title} {Out-of-equilibrium electrons and the {H}all
  conductance of a {F}loquet topological insulator},\ }\href
  {https://doi.org/10.1103/PhysRevB.91.155422} {\bibfield  {journal} {\bibinfo
  {journal} {Phys. Rev. B}\ }\textbf {\bibinfo {volume} {91}},\ \bibinfo
  {pages} {155422} (\bibinfo {year} {2015})}\BibitemShut {NoStop}%
\bibitem [{\citenamefont {Hsieh}\ \emph {et~al.}(2009)\citenamefont {Hsieh},
  \citenamefont {Xia}, \citenamefont {Wray}, \citenamefont {Qian},
  \citenamefont {Pal}, \citenamefont {Dil}, \citenamefont {Osterwalder},
  \citenamefont {Meier}, \citenamefont {Bihlmayer}, \citenamefont {Kane},
  \citenamefont {Hor}, \citenamefont {Cava},\ and\ \citenamefont
  {Hasan}}]{doi:10.1126/science.1167733}%
  \BibitemOpen
  \bibfield  {author} {\bibinfo {author} {\bibfnamefont {D.}~\bibnamefont
  {Hsieh}}, \bibinfo {author} {\bibfnamefont {Y.}~\bibnamefont {Xia}}, \bibinfo
  {author} {\bibfnamefont {L.}~\bibnamefont {Wray}}, \bibinfo {author}
  {\bibfnamefont {D.}~\bibnamefont {Qian}}, \bibinfo {author} {\bibfnamefont
  {A.}~\bibnamefont {Pal}}, \bibinfo {author} {\bibfnamefont {J.~H.}\
  \bibnamefont {Dil}}, \bibinfo {author} {\bibfnamefont {J.}~\bibnamefont
  {Osterwalder}}, \bibinfo {author} {\bibfnamefont {F.}~\bibnamefont {Meier}},
  \bibinfo {author} {\bibfnamefont {G.}~\bibnamefont {Bihlmayer}}, \bibinfo
  {author} {\bibfnamefont {C.~L.}\ \bibnamefont {Kane}}, \bibinfo {author}
  {\bibfnamefont {Y.~S.}\ \bibnamefont {Hor}}, \bibinfo {author} {\bibfnamefont
  {R.~J.}\ \bibnamefont {Cava}},\ and\ \bibinfo {author} {\bibfnamefont
  {M.~Z.}\ \bibnamefont {Hasan}},\ }\bibfield  {title} {\bibinfo {title}
  {Observation of unconventional quantum spin textures in topological
  insulators},\ }\href {https://doi.org/10.1126/science.1167733} {\bibfield
  {journal} {\bibinfo  {journal} {Science}\ }\textbf {\bibinfo {volume}
  {323}},\ \bibinfo {pages} {919} (\bibinfo {year} {2009})}\BibitemShut
  {NoStop}%
\bibitem [{\citenamefont {Iwasawa}\ \emph {et~al.}(2024)\citenamefont
  {Iwasawa}, \citenamefont {Ueno}, \citenamefont {Iwata}, \citenamefont
  {Kuroda}, \citenamefont {Kokh}, \citenamefont {Tereshchenko}, \citenamefont
  {Miyamoto}, \citenamefont {Kimura},\ and\ \citenamefont
  {Okuda}}]{Iwasawa2024}%
  \BibitemOpen
  \bibfield  {author} {\bibinfo {author} {\bibfnamefont {H.}~\bibnamefont
  {Iwasawa}}, \bibinfo {author} {\bibfnamefont {T.}~\bibnamefont {Ueno}},
  \bibinfo {author} {\bibfnamefont {T.}~\bibnamefont {Iwata}}, \bibinfo
  {author} {\bibfnamefont {K.}~\bibnamefont {Kuroda}}, \bibinfo {author}
  {\bibfnamefont {K.~A.}\ \bibnamefont {Kokh}}, \bibinfo {author}
  {\bibfnamefont {O.~E.}\ \bibnamefont {Tereshchenko}}, \bibinfo {author}
  {\bibfnamefont {K.}~\bibnamefont {Miyamoto}}, \bibinfo {author}
  {\bibfnamefont {A.}~\bibnamefont {Kimura}},\ and\ \bibinfo {author}
  {\bibfnamefont {T.}~\bibnamefont {Okuda}},\ }\bibfield  {title} {\bibinfo
  {title} {Efficiency improvement of spin-resolved {ARPES} experiments using
  {G}aussian process regression},\ }\href
  {https://doi.org/10.1038/s41598-024-66704-8} {\bibfield  {journal} {\bibinfo
  {journal} {Scientific Reports}\ }\textbf {\bibinfo {volume} {14}},\ \bibinfo
  {pages} {20970} (\bibinfo {year} {2024})}\BibitemShut {NoStop}%
\bibitem [{\citenamefont {Sentef}\ \emph {et~al.}(2015)\citenamefont {Sentef},
  \citenamefont {Claassen}, \citenamefont {Kemper}, \citenamefont {Moritz},
  \citenamefont {Oka}, \citenamefont {Freericks},\ and\ \citenamefont
  {Devereaux}}]{SentefTrARPES}%
  \BibitemOpen
  \bibfield  {author} {\bibinfo {author} {\bibfnamefont {M.}~\bibnamefont
  {Sentef}}, \bibinfo {author} {\bibfnamefont {M.}~\bibnamefont {Claassen}},
  \bibinfo {author} {\bibfnamefont {A.}~\bibnamefont {Kemper}}, \bibinfo
  {author} {\bibfnamefont {B.}~\bibnamefont {Moritz}}, \bibinfo {author}
  {\bibfnamefont {T.}~\bibnamefont {Oka}}, \bibinfo {author} {\bibfnamefont
  {J.}~\bibnamefont {Freericks}},\ and\ \bibinfo {author} {\bibfnamefont
  {T.}~\bibnamefont {Devereaux}},\ }\bibfield  {title} {\bibinfo {title}
  {Theory of {Floquet} band formation and local pseudospin textures in
  pump-probe photoemission of graphene},\ }\href
  {https://doi.org/10.1038/ncomms8047} {\bibfield  {journal} {\bibinfo
  {journal} {Nature Communications}\ }\textbf {\bibinfo {volume} {6}},\
  \bibinfo {pages} {7047} (\bibinfo {year} {2015})}\BibitemShut {NoStop}%
\bibitem [{\citenamefont {Freericks}\ \emph {et~al.}(2009)\citenamefont
  {Freericks}, \citenamefont {Krishnamurthy},\ and\ \citenamefont
  {Pruschke}}]{FreeeicksTrARPES}%
  \BibitemOpen
  \bibfield  {author} {\bibinfo {author} {\bibfnamefont {J.~K.}\ \bibnamefont
  {Freericks}}, \bibinfo {author} {\bibfnamefont {H.~R.}\ \bibnamefont
  {Krishnamurthy}},\ and\ \bibinfo {author} {\bibfnamefont {T.}~\bibnamefont
  {Pruschke}},\ }\bibfield  {title} {\bibinfo {title} {Theoretical description
  of time-resolved photoemission spectroscopy: Application to pump-probe
  experiments},\ }\href {https://doi.org/10.1103/PhysRevLett.102.136401}
  {\bibfield  {journal} {\bibinfo  {journal} {Phys. Rev. Lett.}\ }\textbf
  {\bibinfo {volume} {102}},\ \bibinfo {pages} {136401} (\bibinfo {year}
  {2009})}\BibitemShut {NoStop}%
\bibitem [{\citenamefont {Farrell}\ \emph {et~al.}(2016)\citenamefont
  {Farrell}, \citenamefont {Arsenault},\ and\ \citenamefont
  {Pereg-Barnea}}]{FarrellTrARPES}%
  \BibitemOpen
  \bibfield  {author} {\bibinfo {author} {\bibfnamefont {A.}~\bibnamefont
  {Farrell}}, \bibinfo {author} {\bibfnamefont {A.}~\bibnamefont {Arsenault}},\
  and\ \bibinfo {author} {\bibfnamefont {T.}~\bibnamefont {Pereg-Barnea}},\
  }\bibfield  {title} {\bibinfo {title} {Dirac cones, {F}loquet side bands, and
  theory of time-resolved angle-resolved photoemission},\ }\href
  {https://doi.org/10.1103/PhysRevB.94.155304} {\bibfield  {journal} {\bibinfo
  {journal} {Phys. Rev. B}\ }\textbf {\bibinfo {volume} {94}},\ \bibinfo
  {pages} {155304} (\bibinfo {year} {2016})}\BibitemShut {NoStop}%
\end{thebibliography}%
				
		{\allowdisplaybreaks
        \onecolumngrid
        \clearpage
        \begin{center}
        \textbf{\large Supplemental Materials for ``{\zb Parity-selective spin splitting in coplanar antiferromagnets via bichromatic driving}''}\\
        \vspace{4mm}
	     {Di Zhu,$^{1}$ Zhongbo Yan,$^{1}$ and Mohsen Yarmohammadi$^{2}$}\\
         \vspace{2mm}
	{\em $^1${Guangdong Provincial Key Laboratory of Magnetoelectric Physics and Devices, State Key Laboratory of Optoelectronic Materials and Technologies, School of Physics, Sun Yat-sen University, Guangzhou 510275, China}}\\
    {\em $^2$ Department of Physics, Georgetown University, Washington DC 20057, USA}
        \end{center}
        
        \setcounter{equation}{0}
		
		\setcounter{equation}{0}
		\renewcommand{\theequation}{S\arabic{equation}}
		\setcounter{figure}{0}
		\renewcommand{\thefigure}{S\arabic{figure}}
		\setcounter{section}{0}
\renewcommand{\thesection}{S\arabic{section}}
\setcounter{subsection}{0}
\renewcommand{\thesubsection}{\Alph{subsection}}
\setcounter{secnumdepth}{2}
		\setcounter{table}{0}
		\renewcommand{\thetable}{S\arabic{table}}

        \tableofcontents
		
		\section{Derivation of the effective Hamiltonian for the bichromatically ($\omega$--$2\omega$) driven coplanar antiferromagnet}
		
		We start from the lattice Hamiltonian of the all-out BCAFM, \begin{align}
			\mathcal{H}(\bm{k}) = 2t h_x^c \sigma_x + 2t h_y^c \tau_x + 4t_s h_x^c h_y^c \tau_x \sigma_x 
			+ 4t_a h_x^c h_y^c \tau_y \sigma_y - M_x \sigma_z s_x + M_y \tau_z s_y,
		\end{align}where $h_{x,y}^c=\cos k_{x,y}$. To facilitate an analytical derivation of the Floquet effective Hamiltonian, we expand the lattice model around small momenta. Keeping terms up to second order in $k_x$ and $k_y$ gives
        \begin{align}
			\mathcal{H}(\bm{k})= {}& 2t \left(1-k_x^2/2\right) \sigma_x + 2t \left(1-k_y^2/2\right) \tau_x + 4t_s \left(1-k_x^2/2\right) \left(1-k_y^2/2\right) \tau_x \sigma_x \notag \\{}
			&+ 4t_a \left(1-k_x^2/2\right) \left(1-k_y^2/2\right)\tau_y \sigma_y - M_x \sigma_z s_x + M_y \tau_z s_y.
		\end{align}
        In the following, we neglect the higher-order cross term proportional to $k_x^2k_y^2$.
		
		\subsection{Effective Floquet Hamiltonian for the ($\omega$--$2\omega$) BLPL}
		
		For bichromatic linearly polarized light (BLPL), the vector potential is taken as
		\begin{subequations}
			\begin{align}
				A_x(t) &= A_0\left[\cos\omega t+S\cos(2\omega t+\phi)\cos\beta\right], \\
				A_y(t) &= A_0S\cos(2\omega t+\phi)\sin\beta.
			\end{align}
		\end{subequations}Setting $e=\hbar=1$, we couple the electromagnetic field to the electrons via the Peierls substitution, $k_x \rightarrow k_x+A_x(t)$ and $k_y \rightarrow k_y+A_y(t)$. Expanding to second order in momentum, the diagonal kinetic terms transform as
		\begin{subequations}
			\begin{align}
				1-\frac{k_x^2}{2} &\rightarrow 1-\frac{(k_x+A_x)^2}{2} = 1-\frac{k_x^2}{2}-k_xA_x-\frac{A_x^2}{2}, \\
				1-\frac{k_y^2}{2} &\rightarrow 1-\frac{(k_y+A_y)^2}{2} = 1-\frac{k_y^2}{2}-k_yA_y-\frac{A_y^2}{2}.
			\end{align}
		\end{subequations}Similarly, the mixed kinetic factor becomes
		\begin{equation}
			1-\frac{k_x^2}{2}-\frac{k_y^2}{2} \rightarrow 1-\frac{(k_x+A_x)^2}{2}-\frac{(k_y+A_y)^2}{2}.
		\end{equation}
		To simplify the notation, we introduce the trigonometric shorthands $c_\beta=\cos\beta$ and $s_\beta=\sin\beta$. The time-dependent Hamiltonian can then be compactly written as
		\begin{equation}
			\begin{aligned}
				\mathcal{H}(\bm{k}+\bm{A}(t)) ={} 2tX(t)\sigma_x + 2tY(t)\tau_x + 4t_sR(t)\tau_x\sigma_x + 4t_aR(t)\tau_y\sigma_y - M_x\sigma_zs_x + M_y\tau_zs_y,
			\end{aligned}
		\end{equation}
		where the time-periodic kinematic factors are defined as
		\begin{subequations}
			\begin{align}
				X(t) &= 1-\frac{k_x^2}{2}-k_xA_x(t)-\frac{A_x^2(t)}{2}, \\
				Y(t) &= 1-\frac{k_y^2}{2}-k_yA_y(t)-\frac{A_y^2(t)}{2}, \\
				R(t) &= 1-\frac{k_x^2}{2}-\frac{k_y^2}{2}-k_xA_x(t)-k_yA_y(t)-\frac{A_x^2(t)+A_y^2(t)}{2}.
			\end{align}
		\end{subequations}
		
		We expand the time-periodic Hamiltonian in terms of its Floquet harmonics,
		\begin{subequations}
			\begin{align}
				\mathcal{H}_m &= \frac{1}{T}\int_0^T \mathcal{H}(\bm{k}+\bm{A}(t))e^{-im\omega t}\,dt, \\
				\mathcal{H}(\bm{k}+\bm{A}(t)) &= \sum_m H_me^{im\omega t}.
			\end{align}
		\end{subequations}
		Equivalently, the Fourier coefficients of $X(t)$, $Y(t)$, and $R(t)$ are denoted by $X_m$, $Y_m$, and $R_m$, respectively. The static coefficients are
		\begin{subequations}
			\begin{align}
				X_0 &= 1-\frac{k_x^2}{2}-\frac{A_0^2}{4}\left(1+S^2c_\beta^2\right), \\
				Y_0 &= 1-\frac{k_y^2}{2}-\frac{A_0^2S^2s_\beta^2}{4}, \\
				R_0 &= 1-\frac{k_x^2}{2}-\frac{k_y^2}{2}-\frac{A_0^2}{4}\left(1+S^2\right).
			\end{align}
		\end{subequations}
		For the first harmonic, one obtains
		\begin{subequations}
			\begin{align}
				X_{\pm1} &= -\frac{A_0k_x}{2}-\frac{A_0^2S c_\beta}{4}e^{\pm i\phi}, \\
				Y_{\pm1} &= 0, \\
				R_{\pm1} &= -\frac{A_0k_x}{2}-\frac{A_0^2S c_\beta}{4}e^{\pm i\phi}.
			\end{align}
		\end{subequations}
		For the second harmonic, the coefficients are
		\begin{subequations}
			\begin{align}
				X_{\pm2} &= -\frac{A_0S k_x c_\beta}{2}e^{\pm i\phi}-\frac{A_0^2}{8}, \\
				Y_{\pm2} &= -\frac{A_0S k_y s_\beta}{2}e^{\pm i\phi}, \\
				R_{\pm2} &= -\frac{A_0S}{2}\left(k_xc_\beta+k_ys_\beta\right)e^{\pm i\phi}-\frac{A_0^2}{8}.
			\end{align}
		\end{subequations}
		The non-static Floquet harmonics of the Hamiltonian ($m \neq 0$) are therefore
		\begin{equation}
			\mathcal{H}_m = 2tX_{m}\sigma_x + 2tY_{m}\tau_x + 4t_sR_{m}\tau_x\sigma_x + 4t_aR_{m}\tau_y\sigma_y,
		\end{equation}
		and the static component is
		\begin{equation}
			\begin{aligned}
				\mathcal{H}_0 ={} 2tX_0\sigma_x + 2tY_0\tau_x + 4t_sR_0\tau_x\sigma_x + 4t_aR_0\tau_y\sigma_y - M_x\sigma_zs_x + M_y\tau_zs_y.
			\end{aligned}
		\end{equation}
		
		Using the high-frequency Floquet formalism, the effective Hamiltonian expanded to order $\mathcal{O}(\omega^{-1})$ is given by
		\begin{equation}
			\mathcal{H}_{\rm eff}(\bm{k}) \approx \mathcal{H}_0(\bm{k}) + \frac{[\mathcal{H}_1(\bm{k}),\mathcal{H}_{-1}(\bm{k})]}{\omega} + \frac{[\mathcal{H}_2(\bm{k}),\mathcal{H}_{-2}(\bm{k})]}{2\omega}.
		\end{equation}
		Evaluating the relevant commutators yields
		\begin{subequations}
			\begin{align}
				[\sigma_x,\tau_y\sigma_y] &= 2i\tau_y\sigma_z, \\
				[\tau_x,\tau_y\sigma_y] &= 2i\tau_z\sigma_y.
			\end{align}
		\end{subequations}
		We note that the $t_s$ term commutes with all other relevant components of the Hamiltonian,
		\begin{equation}
			[\tau_x\sigma_x,\sigma_x] = [\tau_x\sigma_x,\tau_x] = [\tau_x\sigma_x,\tau_y\sigma_y] = 0,
		\end{equation}
		and thus does not contribute to the commutator corrections. For $m=1$, since $X_{\pm1}=R_{\pm1}$ and $Y_{\pm1}=0$, the first-order correction vanishes identically:
		\begin{equation}
			[\mathcal{H}_1(\bm{k}),\mathcal{H}_{-1}(\bm{k})] = 0.
		\end{equation}
		For $m=2$, the commutator evaluates to
		\begin{equation}
			[\mathcal{H}_2(\bm{k}),\mathcal{H}_{-2}(\bm{k})] = 2tt_aA_0^3S k_y\sin\beta\sin\phi \left(\sigma_z\tau_y-\tau_z\sigma_y\right).
		\end{equation}
		Combining these results, the full effective Floquet Hamiltonian is
		\begin{align}
				\mathcal{H}^{\rm BLPL}_{\rm eff}(\bm{k}) ={}& 2t\left[1-\frac{k_x^2}{2}-\frac{A_0^2}{4}\left(1+S^2\cos^2\beta\right)\right]\sigma_x + 2t\left[1-\frac{k_y^2}{2}-\frac{A_0^2S^2\sin^2\beta}{4}\right]\tau_x \notag\\
				&+ 4t_s\left[1-\frac{k_x^2}{2}-\frac{k_y^2}{2}-\frac{A_0^2}{4}\left(1+S^2\right)\right]\tau_x\sigma_x + 4t_a\left[1-\frac{k_x^2}{2}-\frac{k_y^2}{2}-\frac{A_0^2}{4}\left(1+S^2\right)\right]\tau_y\sigma_y \notag\\
				&- M_x\sigma_zs_x + M_y\tau_zs_y+ \frac{tt_aA_0^3S}{\omega} k_y\sin\beta\sin\phi \left(\tau_y\sigma_z-\tau_z\sigma_y\right).
		\end{align}
		
		\subsection{Effective Floquet Hamiltonian for the ($\omega$--$2\omega$) BCPL}
		
		For BCPL, the vector potential is
		\begin{subequations}\label{eq:2omegaCPL}
			\begin{align}
				A_x(t) &= A_0[ \cos(\omega t) + S \cos(2\omega t + \phi)], \\
				A_y(t) &= A_0[\eta_1\sin(\omega t) + \eta_2 S \sin(2\omega t + \phi)].
			\end{align}
		\end{subequations}Applying the Peierls substitution gives
       \begin{equation}
			\begin{aligned}
				\mathcal{H}(\bm{k}+\bm{A}(t)) =2tX(t)\sigma_x + 2tY(t)\tau_x + 4t_sR(t)\tau_x\sigma_x + 4t_aR(t)\tau_y\sigma_y - M_x\sigma_zs_x + M_y\tau_zs_y,
			\end{aligned}
		\end{equation}
        with
		\begin{subequations}
			\begin{align}
				X(t) &= 1-\frac{k_x^2}{2}-k_xA_x(t)-\frac{A_x^2(t)}{2}, \\
				Y(t) &= 1-\frac{k_y^2}{2}-k_yA_y(t)-\frac{A_y^2(t)}{2}, \\
				R(t) &= 1-\frac{k_x^2}{2}-\frac{k_y^2}{2}-k_xA_x(t)-k_yA_y(t)-\frac{A_x^2(t)+A_y^2(t)}{2}.
			\end{align}
		\end{subequations}
        
        The static Fourier coefficients are
        \begin{subequations}
			\begin{align}
				X_0 &= 1-\frac{k_x^2}{2}-\frac{A_0^2}{4}(1+S^2), \\
				Y_0 &= 1-\frac{k_y^2}{2}-\frac{A_0^2}{4}(1+S^2), \\
				R_0 &= 1-\frac{k_x^2}{2}-\frac{k_y^2}{2}-\frac{A_0^2}{2}(1+S^2).
			\end{align}
		\end{subequations}
        For the first harmonic, the coefficients evaluate to
        \begin{subequations}
			\begin{align}
				X_{\pm1} &= -\frac{A_0}{2}k_x-\frac{A_0^2}{4}Se^{\pm i\phi}, \\
				Y_{\pm1} &= -\eta_1\frac{A_0}{2i}k_y-\eta_1\eta_2\frac{A_0^2}{4}Se^{\pm i\phi}, \\
				R_{\pm1} &= -\frac{A_0}{2}k_x-\frac{A_0^2}{4}Se^{\pm i\phi}-\eta_1\frac{A_0}{2i}k_y-\eta_1\eta_2\frac{A_0^2}{4}Se^{\pm i\phi}.
			\end{align}
		\end{subequations}
        For the second harmonic, one obtains
        \begin{subequations}
			\begin{align}
				X_{\pm2} &=-\frac{A_0}{2}Sk_xe^{\pm i\phi}-\frac{A_0^2}{8}, \\
				Y_{\pm2} &= -\frac{A_0}{2i}\eta_2 Sk_ye^{\pm i\phi}+\frac{A_0^2}{8}, \\
				R_{\pm2} &= -\frac{A_0}{2}Sk_xe^{\pm i\phi}-\frac{A_0}{2i}\eta_2 Sk_ye^{\pm i\phi}.
			\end{align}
		\end{subequations}
        
        Using the high-frequency expansion,
		\begin{equation}
			\mathcal{H}_{\rm eff}(\bm{k}) \approx \mathcal{H}_0(\bm{k}) + \frac{[\mathcal{H}_1(\bm{k}),\mathcal{H}_{-1}(\bm{k})]}{\omega} + \frac{[\mathcal{H}_2(\bm{k}),\mathcal{H}_{-2}(\bm{k})]}{2\omega},
        \end{equation}
        the relevant commutators are
        \begin{equation}
        \begin{split}[\mathcal{H}_{1},\mathcal{H}_{-1}]=&4tt_a\left[{A_0^3}\eta_1\eta_2 Sk_x\sin\phi - 2\eta_1A_0^2k_xk_y - {A_0^3}S\eta_1k_y\cos\phi\right]\tau_y\sigma_z \\
        &+ 4tt_a\left[{2\eta_1A_0^2}k_xk_y - {A_0^3}\eta_1\eta_2 Sk_x\sin\phi + {A_0^3}S\eta_1k_y\cos\phi\right]\tau_z\sigma_y,
        \end{split}
        \end{equation}
        and 
        \begin{equation}
        \begin{split}[\mathcal{H}_{2},\mathcal{H}_{-2}]=&2tt_a\left({A_0^3}Sk_x\sin\phi - {4A_0^2}\eta_2 S^2k_xk_y - {A_0^3}\eta_2 Sk_y\cos\phi\right)\tau_y\sigma_z\\ 
        &+ 2tt_a\left({4A_0^2}\eta_2 S^2k_xk_y - {A_0^3}Sk_x\sin\phi + {A_0^3}\eta_2 Sk_y\cos\phi\right)\tau_z\sigma_y.
        \end{split}
        \end{equation}
        
		Combining these terms, the effective Floquet Hamiltonian is
		\begin{equation}
        \begin{split}
				\mathcal{H}^{\rm BCPL}_{\rm eff}(\bm{k}) ={}& 2t \left[1-\frac{k_x^2}{2}-\frac{A_0^2}{4}(1+S^2)\right] \sigma_x + 2t \left[1-\frac{k_y^2}{2}-\frac{A_0^2}{4}(1+S^2)\right] \tau_x \\
				&+ 4t_s \left[1-\frac{k_x^2}{2}-\frac{k_y^2}{2}-\frac{A_0^2}{2}(1+S^2)\right] \tau_x \sigma_x + 4t_a \left[1-\frac{k_x^2}{2}-\frac{k_y^2}{2}-\frac{A_0^2}{2}(1+S^2)\right]\tau_y \sigma_y\\
                &- M_x \sigma_z s_x + M_y \tau_z s_y \\
				&+ \frac{tt_a}{\omega}\left[(4\eta_1\eta_2+1)SA_0^3k_x\sin\phi - (8\eta_1+4\eta_2 S^2)A_0^2k_xk_y - (4\eta_1+\eta_2)SA_0^3k_y\cos\phi\right]\tau_y\sigma_z \\
				&+ \frac{tt_a}{\omega}\left[(8\eta_1+4\eta_2 S^2)A_0^2k_xk_y - (4\eta_1\eta_2+1)SA_0^3k_x\sin\phi + (4\eta_1+\eta_2)SA_0^3k_y\cos\phi\right]\tau_z\sigma_y.
		\end{split}
        \end{equation}

		\subsection{Effective Floquet Hamiltonian for the ($\omega$--$2\omega$) BCLPL}
		
		For the hybrid BCLPL configuration, the vector potential is

		\begin{subequations}\label{eq:bclpl_A}
			\begin{align}
				A_x(t) &= A_0\left[\cos\omega t+S\cos(2\omega t+\phi)\cos\beta\right], \\
				A_y(t) &= A_0\left[\eta_1\sin\omega t+S\cos(2\omega t+\phi)\sin\beta\right],
			\end{align}
		\end{subequations}
		where $\eta_1=\pm1$ denotes the helicity of the fundamental circularly polarized beam. Using $c_\beta=\cos\beta$ and $s_\beta=\sin\beta$, the time-dependent Hamiltonian takes the form
		\begin{equation}
			\begin{aligned}
				\mathcal{H}(\bm{k}+\bm{A}(t)) ={}& 2tX(t)\sigma_x + 2tY(t)\tau_x + 4t_sR(t)\tau_x\sigma_x + 4t_aR(t)\tau_y\sigma_y - M_x\sigma_zs_x + M_y\tau_zs_y,
			\end{aligned}
		\end{equation}
		with
		\begin{subequations}
			\begin{align}
				X(t) &= 1-\frac{k_x^2}{2}-k_xA_x-\frac{A_x^2}{2}, \\
				Y(t) &= 1-\frac{k_y^2}{2}-k_yA_y-\frac{A_y^2}{2}, \\
				R(t) &= 1-\frac{k_x^2}{2}-\frac{k_y^2}{2}-k_xA_x-k_yA_y-\frac{A_x^2+A_y^2}{2}.
			\end{align}
		\end{subequations}
        
        The static Fourier coefficients are
		\begin{subequations}
			\begin{align}
				X_0 &= 1-\frac{k_x^2}{2}-\frac{A_0^2}{4}\left(1+S^2c_\beta^2\right), \\
				Y_0 &= 1-\frac{k_y^2}{2}-\frac{A_0^2}{4}\left(\eta_1^2+S^2s_\beta^2\right), \\
				R_0 &= 1-\frac{k_x^2}{2}-\frac{k_y^2}{2}-\frac{A_0^2}{4}\left(1+\eta_1^2+S^2\right).
			\end{align}
		\end{subequations}
		For the first harmonic, we obtain
		\begin{subequations}
			\begin{align}
				X_{\pm1} &= -\frac{A_0k_x}{2}-\frac{A_0^2S c_\beta}{4}e^{\pm i\phi}, \\
				Y_{\pm1} &= \pm\frac{iA_0\eta_1 k_y}{2}\mp\frac{iA_0^2\eta_1 S s_\beta}{4}e^{\pm i\phi}, \\
				R_{\pm1} &= X_{\pm1}+Y_{\pm1}.
			\end{align}
		\end{subequations}
		For the second harmonic, the coefficients are
		\begin{subequations}
			\begin{align}
				X_{\pm2} &= -\frac{A_0S k_x c_\beta}{2}e^{\pm i\phi}-\frac{A_0^2}{8}, \\
				Y_{\pm2} &= -\frac{A_0S k_y s_\beta}{2}e^{\pm i\phi}+\frac{A_0^2\eta_1^2}{8}, \\
				R_{\pm2} &= X_{\pm2}+Y_{\pm2}.
			\end{align}
		\end{subequations}
        For the third harmonic, we have
        \begin{subequations}
			\begin{align}
				X_{\pm3} &= -\frac{A_0^2Sc_\beta}{4}e^{\pm i\phi}, \\
				Y_{\pm3} &= \pm\frac{i\eta_1A_0^2Ss_\beta}{4}e^{\pm i\phi}, \\
				R_{\pm3} &= X_{\pm3}+Y_{\pm3}.
			\end{align}
		\end{subequations}
		
		For $m\neq0$, the Floquet harmonics are
		\begin{equation}
			\mathcal{H}_m = 2tX_{m}\sigma_x + 2tY_{m}\tau_x + 4t_sR_{m}\tau_x\sigma_x + 4t_aR_{m}\tau_y\sigma_y,
		\end{equation}
		while the static component is
		\begin{align}
			\mathcal{H}_0 ={} 2tX_0\sigma_x+2tY_0\tau_x + 4t_sR_0\tau_x\sigma_x + 4t_aR_0\tau_y\sigma_y- M_x\sigma_zs_x + M_y\tau_zs_y.
		\end{align}
        
		The relevant commutators are
		\begin{equation}
				[\mathcal{H}_{1},\mathcal{H}_{-1}] = -16tt_a\eta_1\left[\frac{A_0^2}{2}k_xk_y+\frac{A_0^3S}{4}\left(k_y\cos\beta-k_x\sin\beta\right)\cos\phi-\frac{A_0^4S^2}{8}\cos\beta\sin\beta\right]\left(\tau_y\sigma_z-\tau_z\sigma_y\right),
		\end{equation}
        and 
        \begin{equation}
				[\mathcal{H}_{2},\mathcal{H}_{-2}] = 2tt_aA_0^3S\left(k_y\sin\beta+\eta_1^2k_x\cos\beta\right)\sin\phi\left(\tau_y\sigma_z-\tau_z\sigma_y\right),
		\end{equation}
        and
        \begin{equation}
				[\mathcal{H}_{3},\mathcal{H}_{-3}] = -2tt_a\eta_1A_0^4S\cos\beta\sin\beta\left(\tau_y\sigma_z-\tau_z\sigma_y\right).
		\end{equation}
		
		Using $\eta_1^2=1$, the effective Floquet Hamiltonian up to order $\mathcal{O}(\omega^{-1})$ is
		\begin{equation}
			\begin{aligned}
				\mathcal{H}_{\rm eff}^{\rm BCLPL}(\bm{k}) ={}& 2t\left[1-\frac{k_x^2}{2}-\frac{A_0^2}{4}\left(1+S^2\cos^2\beta\right)\right]\sigma_x + 2t\left[1-\frac{k_y^2}{2}-\frac{A_0^2}{4}\left(1+S^2\sin^2\beta\right)\right]\tau_x \\
				&+ 4t_s\left[1-\frac{k_x^2}{2}-\frac{k_y^2}{2}-\frac{A_0^2}{4}\left(2+S^2\right)\right]\tau_x\sigma_x + 4t_a\left[1-\frac{k_x^2}{2}-\frac{k_y^2}{2}-\frac{A_0^2}{4}\left(2+S^2\right)\right]\tau_y\sigma_y \\
				&- M_x\sigma_zs_x + M_y\tau_zs_y \\
				&- \frac{tt_a}{\omega}\left[{8A_0^2\eta_1}k_xk_y+{A_0^3S}k_y\left(4\eta_1\cos\beta\cos\phi-\sin\beta\sin\phi\right)\right]\left(\tau_y\sigma_z-\tau_z\sigma_y\right)\\
                &+ \frac{tt_a}{\omega}\left[{A_0^3S}k_x\left(4\eta_1\sin\beta\cos\phi+\cos\beta\sin\phi\right)+\frac{4}{3}{A_0^4S^2}\eta_1\cos\beta\sin\beta\right]\left(\tau_y\sigma_z-\tau_z\sigma_y\right).
			\end{aligned}
		\end{equation}
		
		In the preceding derivations, we truncated the high-frequency expansion at $m = 2$. However, accurately capturing the induced terms under $\omega$--$n\omega$ driving requires extending this expansion, $\sum_m [H_{-m}, H_{m}] / (m\hbar\omega)$, up to order $m=2n$. In the BLPL and BCPL protocols, terms with $m \ge 3$ do not contribute because their corresponding high-frequency commutators, $[H_{-m}, H_m]$, evaluate to zero. In BCLPL protocol, although $m=4$ harmonics also emerge from the purely squared $2\omega$ fields, their commutators intrinsically vanish ($[H_{-4}, H_4] = 0$), and all $m > 4$ harmonics are kinematically forbidden. Thus, for BCLPL, the expansion exacts finite corrections at $m=3$, while all $m \ge 4$ terms rigorously evaluate to zero. For brevity, we summarize the final analytical expressions in the following tables.
		
		\section{High-temperature expansion of the spin polarization}
        To analytically capture the out-of-plane spin polarization in our all-out BCAFM, we evaluate it with the high-temperature expansion~\cite{Hayami2020AM1,Hayami2020AM2}\begin{align}
            {\rm Tr}[e^{\beta {\zb \mathcal{H}}_{\rm eff}(\bm{k})}s_z]=\sum_s\frac{(-\beta)^s}{s!}g_s^z(\bm{k})\,,
        \end{align}where $\beta$ is the inverse temperature and\begin{align}g_s^z(\bm{k}) = {\rm Tr}[{\zb \mathcal{H}}_{\rm eff}^s(\bm{k})s_z]\,.\end{align}Expressing the general Floquet effective Hamiltonian for various driving configurations $\alpha \in\{\rm BLPL, BCPL, BCLPL\}$ as\begin{align}
{\zb \mathcal{H}}_{\rm eff}(\bm{k})={} &2tX^\alpha\sigma_x+2tY^\alpha\tau_x+4t_sR^\alpha\sigma_x\tau_x+4t_aR^\alpha\sigma_y\tau_y-M_x\sigma_zs_x+M_y\tau_zs_y+F^\alpha(\tau_y\sigma_z-\tau_z\sigma_y)\,,
\end{align}we find that the lowest nonvanishing contribution to the spin polarization arises at fifth order:
\begin{equation}
g_5^z(\bm{k})\propto tt_s(X^\alpha+Y^\alpha)R^\alpha F^\alpha M_xM_y\,.
\end{equation}
Because $(A^\alpha+B^\alpha)$ and $C^\alpha$ transform as $s$-wave terms, the spatial parity of the emergent spin texture is strictly determined by $F^\alpha$. Consequently, our discussion of parity control fundamentally reduces to analyzing the detailed symmetries of the $F^\alpha$ coefficient.
		
		\section{Floquet effective Hamiltonians for bichromatic driving: Fundamental vs. higher harmonics}
		
		In this section, we provide the explicit analytical expressions for the coefficients of the effective Floquet Hamiltonian terms corresponding to a coplanar antiferromagnet driven by bichromatic fields. The Tabs.~\ref{tabs1}-\ref{tabs3} below detail the dynamically generated terms resulting from an $\omega$--$n\omega$ driving scheme, comparing the outcomes for the second harmonic ($n=2$) and higher-order harmonics ($n \geq 3$ or $n \geq 4$). To systematically analyze the role of light polarization, we categorize our results into three distinct bichromatic driving configurations: (i) bichromatic linearly polarized light (BLPL): both the fundamental and harmonic beams are linearly polarized, (ii) bichromatic circularly polarized light (BCPL): both beams are circularly polarized, and (iii) bichromatic circular-linear polarized light (BCLPL): a hybrid configuration mixing circular and linear polarizations.
		
		A central result of this exact analytical expansion is the explicit momentum dependence of the dynamically generated non-diagonal terms, specifically $\tau_y\sigma_z$ and $\tau_z\sigma_y$. By inspecting their dependence on $k_x$ and $k_y$, we can extract their spatial parity—revealing whether these induced interactions possess strictly odd, even, or mixed parity under the inversion $\bm{k} \rightarrow -\bm{k}$. This parity breaking is highly sensitive to both the choice of harmonic $n$ and the polarization geometry of the driving fields, the summary of which is discussed in Tab.~II in the main text.\begin{table}[t]
			\renewcommand{\arraystretch}{1.6} 
			\begin{ruledtabular}
				\begin{tabular}{c|c|c}
					Term & Harmonic $n=2$ & Harmonic $n \geq 3$ \\
					\colrule
					$\sigma_x $ & \multicolumn{2}{c}{$\frac{t}{2\hbar^2} \left( 4\hbar^2 - 2\hbar^2k_x^2 - A_0^2e^2\left(1 + S^2\right)  + A_0^2S^2e^2\sin^2\beta \right)$} \\\hline
					$\sigma_z s_x$ & \multicolumn{2}{c}{$-M_x$} \\\hline
					$\tau_x$ & \multicolumn{2}{c}{$\frac{t}{2\hbar^2}\left(4\hbar^2 -2\hbar^2 k_y^2 -A_0^2S^2e^2\sin^2\beta\right)$} \\\hline
					$\tau_x\sigma_x$ & \multicolumn{2}{c}{$-\frac{t_s}{\hbar^2} \left(- 4\hbar^2+ 2\hbar^2\left(k_x^2 + k_y^2\right) + A_0^2e^2\left(1 + S^2\right) \right)$} \\\hline
					$\tau_y\sigma_y$ & \multicolumn{2}{c}{$-\frac{t_a}{\hbar^2} \left(- 4\hbar^2+2\hbar^2\left(k_x^2 + k_y^2\right)+ A_0^2e^2\left(1 + S^2\right) \right)$} \\\hline
					$\tau_y\sigma_z$ & $\displaystyle \frac{A_0^3Se^3k_y t_a t \sin\beta\sin\phi}{\hbar^4\omega}$ & $0$ \\\hline
					$\tau_z s_y$ & \multicolumn{2}{c}{$+M_y$} \\\hline
					$\tau_z\sigma_y$ & $\displaystyle -\frac{A_0^3Se^3k_y t_a t\sin\beta\sin\phi}{\hbar^4\omega}$ & $0$ \\
				\end{tabular}
			\end{ruledtabular}
            \caption{Coefficients of the effective Hamiltonian terms for a coplanar antiferromagnet driven by $\omega$--$n\omega$ bichromatic linearly polarized light (BLPL). The columns compare the generated terms for harmonics $n=2$ and $n \geq 3$, where $A_0$ is the vector potential amplitude of the fundamental $\omega$ beam and $S A_0$ is the amplitude of the $n\omega$ beam.}
			\label{tabs1}
		\end{table}\begin{table*}[t]
			\renewcommand{\arraystretch}{1.9} 
			\begin{ruledtabular}
				\begin{tabular}{c|c|c|c}
					Term & Harmonic $n=2$ & Harmonic $n=3$ & Harmonic $n \geq 4$ \\
					\colrule
					$\sigma_x$ & \multicolumn{3}{c}{$\frac{t}{2\hbar^2} \left(4\hbar^2- 2\hbar^2k_x^2 -A_0^2e^2\left(1 + S^2\right) \right)$} \\\hline
					
					$\sigma_z s_x$ & \multicolumn{3}{c}{$-M_x$} \\\hline
					
					$\tau_x$ & \multicolumn{3}{c}{$\frac{t}{2\hbar^2} \left(4\hbar^2 - 2\hbar^2k_y^2 -A_0^2e^2\left(1 + S^2\right)\right)$} \\\hline
					
					$\tau_x\sigma_x$ & \multicolumn{3}{c}{$-\frac{2t_s}{\hbar^2} \left( - 2\hbar^2+ \hbar^2\left(k_x^2 + k_y^2\right)   + A_0^2e^2\left(1 + S^2\right) \right)$} \\\hline
					
					$\tau_y\sigma_y$ & \multicolumn{3}{c}{$-\frac{2t_a}{\hbar^2} \left(- 2\hbar^2  + \hbar^2\left(k_x^2 + k_y^2\right)   +  A_0^2e^2\left(1 + S^2\right) \right)$} \\\hline
					
					$\tau_y\sigma_z$ & 
					$\begin{aligned}
						&-\frac{A_0^2e^2t_a t}{\hbar^4\omega} \Big( 8\eta_1\hbar k_x k_y - A_0Se\,k_x\sin\phi \\
						&+ 4S^2\eta_2\hbar k_x k_y + 4A_0Se\,\eta_1 k_y\cos\phi \\
						&+ A_0Se\,\eta_2 k_y\cos\phi - 4A_0Se\,\eta_1\eta_2 k_x\sin\phi \Big)
					\end{aligned}$ & 
					$\begin{aligned}
						&-\frac{A_0^2e^2t_a t}{6\hbar^5\omega} \Big( 48\eta_1\hbar^2 k_x k_y \\
						&- 3A_0^2Se^2\sin\phi + 16S^2\eta_2\hbar^2 k_x k_y \\
						&- 3A_0^2Se^2\eta_1\eta_2\sin\phi \Big)
					\end{aligned}$ & 
					$\displaystyle -\frac{8A_0^2e^2k_x k_y t_a t(\eta_2 S^2 + n\eta_1)}{n\hbar^3\omega}$ \\\hline
					
					$\tau_z s_y$ & \multicolumn{3}{c}{$+M_y$} \\\hline
					
					$\tau_z\sigma_y$ & 
					$\begin{aligned}
						&+\frac{A_0^2e^2t_a t}{\hbar^4\omega} \Big( 8\eta_1\hbar k_x k_y - A_0Se\,k_x\sin\phi \\
						&+ 4S^2\eta_2\hbar k_x k_y + 4A_0Se\,\eta_1 k_y\cos\phi \\
						&+ A_0Se\,\eta_2 k_y\cos\phi - 4A_0Se\,\eta_1\eta_2 k_x\sin\phi \Big)
					\end{aligned}$ & 
					$\begin{aligned}
						&+\frac{A_0^2e^2t_a t}{6\hbar^5\omega} \Big( 48\eta_1\hbar^2 k_x k_y \\
						&- 3A_0^2Se^2\sin\phi + 16S^2\eta_2\hbar^2 k_x k_y \\
						&- 3A_0^2Se^2\eta_1\eta_2\sin\phi \Big)
					\end{aligned}$ & 
					$\displaystyle +\frac{8A_0^2e^2k_x k_y t_a t(\eta_2 S^2 + n\eta_1)}{n\hbar^3\omega}$ \\
				\end{tabular}
			\end{ruledtabular}\caption{Coefficients of the effective Hamiltonian terms for a coplanar antiferromagnet driven by $\omega$--$n\omega$ bichromatic circularly polarized light (BCPL). The columns compare the generated terms for harmonics $n=2$, $n=3$, and $n \geq 4$, where $A_0$ is the vector potential amplitude of the fundamental $\omega$ beam and $S A_0$ is the amplitude of the $n\omega$ beam.}
			\label{tabs2}
		\end{table*}\begin{table*}[t]
			\renewcommand{\arraystretch}{1.8} 
			\begin{ruledtabular}
				\begin{tabular}{c|c|c|c}
					Term & Harmonic $n=2$ & Harmonic $n=3$ & Harmonic $n \geq 4$ \\\hline
					\colrule
					$\sigma_x$ & \multicolumn{3}{c}{$\frac{t}{2\hbar^2} \left(4\hbar^2 - 2\hbar^2k_x^2 - A_0^2e^2\left(1 + S^2\right) + A_0^2S^2e^2\sin^2\beta \right)$} \\\hline
					
					$\sigma_z s_x$ & \multicolumn{3}{c}{$-M_x$} \\\hline
					
					$\tau_x$ & \multicolumn{3}{c}{$\frac{t}{2\hbar^2} \left(4\hbar^2 - 2\hbar^2k_y^2 - A_0^2e^2 -  A_0^2S^2e^2\sin^2\beta \right)$} \\\hline
					
					$\tau_x\sigma_x$ & \multicolumn{3}{c}{$-\frac{t_s}{\hbar^2} \left( - 4\hbar^2 + 2\hbar^2\left(k_x^2 + k_y^2\right) + A_0^2e^2\left(2 + S^2\right) \right)$} \\\hline
					
					$\tau_y\sigma_y$ & \multicolumn{3}{c}{$-\frac{t_a}{\hbar^2} \left( - 4\hbar^2 + 2\hbar^2\left(k_x^2 + k_y^2\right) + A_0^2e^2\left(2 + S^2\right)\right)$} \\\hline
					
					$\tau_y\sigma_z$ & 
					$\begin{aligned}
						&+\frac{A_0^2e^2t_a t}{3\hbar^5\omega} \Big( 4A_0^2S^2e^2\eta_1 \cos\beta \sin\beta \\
						&- 24\eta_1\hbar^2k_x k_y + 3A_0Se\hbar k_x \cos\beta \sin\phi \\
						&+ 3A_0Se\hbar k_y \sin\beta \sin\phi \\
						&- 12A_0Se\eta_1\hbar k_y \cos\beta \cos\phi \\
						&+ 12A_0Se\eta_1\hbar k_x \cos\phi \sin\beta \Big)
					\end{aligned}$ & 
					$\begin{aligned}
						&+\frac{A_0^2e^2t_a t}{2\hbar^5\omega} \Big( A_0^2Se^2 \cos\beta \sin\phi \\
						&- 16\eta_1\hbar^2k_x k_y \\
						&+ A_0^2Se^2\eta_1 \cos\phi \sin\beta \\
						&+ A_0^2S^2e^2\eta_1 \cos\beta \sin\beta \Big)
					\end{aligned}$ & 
					$\begin{aligned}
						&-\frac{A_0^2e^2\eta_1 t_a t}{\hbar^5\omega} \Big( -\frac{2}{(n^2-1)} \sin(2\beta) A_0^2S^2e^2 \\
						&+ 8k_x k_y\hbar^2 \Big)
					\end{aligned}$ \\\hline
					
					$\tau_z s_y$ & \multicolumn{3}{c}{$+M_y$} \\\hline
					
					$\tau_z\sigma_y$ & 
					$\begin{aligned}
						&-\frac{A_0^2e^2t_a t}{3\hbar^5\omega} \Big( 4A_0^2S^2e^2\eta_1 \cos\beta \sin\beta \\
						&- 24\eta_1\hbar^2k_x k_y + 3A_0Se\hbar k_x \cos\beta \sin\phi \\
						&+ 3A_0Se\hbar k_y \sin\beta \sin\phi \\
						&- 12A_0Se\eta_1\hbar k_y \cos\beta \cos\phi \\
						&+ 12A_0Se\eta_1\hbar k_x \cos\phi \sin\beta \Big)
					\end{aligned}$ & 
					$\begin{aligned}
						&-\frac{A_0^2e^2t_a t}{2\hbar^5\omega} \Big( A_0^2Se^2 \cos\beta \sin\phi \\
						&- 16\eta_1\hbar^2k_x k_y \\
						&+ A_0^2Se^2\eta_1 \cos\phi \sin\beta \\
						&+ A_0^2S^2e^2\eta_1 \cos\beta \sin\beta \Big)
					\end{aligned}$ & 
					$\begin{aligned}
						&+\frac{A_0^2e^2\eta_1 t_a t}{\hbar^5\omega} \Big( -\frac{2}{(n^2-1)} \sin(2\beta) A_0^2S^2e^2 \\
						&+ 8k_x k_y\hbar^2 \Big)
					\end{aligned}$ \\
				\end{tabular}
			\end{ruledtabular}\caption{Coefficients of the effective Hamiltonian terms for a coplanar antiferromagnet driven by $\omega$--$n\omega$ bichromatic circular-linear polarized light (BCLPL). The columns compare the generated terms for harmonics $n=2$, $n=3$, and $n \geq 4$, where $A_0$ is the vector potential amplitude of the fundamental $\omega$ beam and $S A_0$ is the amplitude of the $n\omega$ beam.}
			\label{tabs3}
		\end{table*}
        
        As discussed before, under $\omega$--$n\omega$ driving, truncating the high-frequency expansion at $m=2$ remains accurate for the BLPL and BCPL protocols, even for $n=2$. In contrast, the BCLPL protocol strictly requires extending the expansion, $\sum_m [H_{-m}, H_{m}] / (m\hbar\omega)$, to order $m=2n$ to accurately capture the induced $s$-wave terms. This requirement arises from the specific structure of the BCLPL vector potential, in which the $y$-component contains a superposition of sine and cosine functions. Because minimal coupling introduces diamagnetic $\bm{A}^2(t)$ terms, the time-dependent Hamiltonian inherently generates higher-order Fourier harmonics. In an $n=2$ drive, truncating the expansion at $m=2$ artificially neglects the destructive quantum interference from $m=3$ virtual photon processes. Table~\ref{tabs3} incorporates these higher-order virtual transitions.
    }
		
	\end{document}